\documentclass[aps,pre,reprint,superscriptaddress]{revtex4-1}

\usepackage{graphicx}
\usepackage{bm}
\usepackage{color}
\usepackage{amsmath}
\usepackage{amssymb}
\usepackage{mathtools}
\usepackage{dsfont}
\usepackage{color}
\usepackage{calc}
\usepackage{hyperref}
\usepackage{multirow}
\usepackage{natmove}

\newcommand{\lla}{\left\langle}
\newcommand{\rra}{\right\rangle}

\graphicspath{{./figures/}}

\begin{document}
\title{Modeling a spheroidal microswimmer and cooperative swimming in thin films}

\author{Mario Theers}
\email{m.theers@fz-juelich.de}
\affiliation{Theoretical Soft Matter and Biophysics, Institute for
Advanced Simulation and Institute of Complex Systems,
Forschungszentrum J\"ulich, D-52425 J\"ulich, Germany}
\author{Elmar Westphal}
\email{e.westphal@fz-juelich.de}
\affiliation{Peter Gr\"unberg Institute and J\"ulich Centre for Neutron
Science, Forschungszentrum J\"ulich, D-52425 J\"ulich, Germany}
\author{Gerhard Gompper}
\email{g.gompper@fz-juelich.de}
\affiliation{Theoretical Soft Matter and Biophysics, Institute for
Advanced Simulation and Institute of Complex Systems,
Forschungszentrum J\"ulich, D-52425 J\"ulich, Germany}
\author{Roland G. Winkler}
\email{r.winkler@fz-juelich.de}
\affiliation{Theoretical Soft Matter and Biophysics, Institute for
Advanced Simulation and Institute of Complex Systems,
Forschungszentrum J\"ulich, D-52425 J\"ulich, Germany}


\date{\today}

\begin{abstract}
We propose a hydrodynamic model for a spheroidal microswimmer with two tangential surface velocity modes. This model is analytically solvable and reduces to Lighthill's and
Blake's spherical squirmer model in the limit of equal major and minor semi-axes.
Furthermore, we present an implementation of such a spheroidal squirmer by means of multiparticle collision dynamics simulations.
We investigate its properties as well as the scattering of two spheroidal squirmers in a slit geometry.
Thereby we find a stable fixed point, where two pullers swim cooperatively forming a wedge-like conformation with a small constant angle.
\end{abstract}
\maketitle

\section{Introduction}

Living matter exhibits a broad spectrum of unique phenomena which emerge as a consequence of its active constituents.   Examples of such systems range from the macroscopic scale of flocks of birds and mammalian herds to the microscopic scale of bacterial suspensions \cite{elge:15,vics:12}. Specifically, active systems exhibit remarkable nonequilibrium phenomena and emergent behavior like swarming \cite{cope:09,darn:10,kear:10,dres:11,part:13}, turbulence \cite{dres:11}, and activity-induced clustering and phase transitions \cite{bial:12,butt:13,mogn:13,theu:12,fily:14.1,yang:14.2,sten:14,fily:14,redn:13,fily:12,gros:12,loba:13,zoet:14,wyso:14}.
The understanding of these collective phenomena requires the characterization of the underlying physical interaction mechanisms. Experiments and simulations indicate that shape-induced interactions, such as inelastic collisions between elongated objects or of active particles with surfaces lead to clustering, collective motion, and surface-induced aggregation \cite{peru:06,dres:11,yang:10,elge:09}. For micrometer-size biological unicellular swimmers, e.g., bacteria ({\em E. coli}), algae ({\em Chlamydomonas}), spermatozoa, or protozoa ({\em Paramecium}), hydrodynamic interactions are considered to be important for collective effects and determine their behavior adjacent to surfaces \cite{berk:08,laug:09,elge:15,laug:06,hu:15,dile:11,leme:13}.

Generic models, which capture the essential swimming aspects, are crucial in theoretical studies of microswimmers. On the one hand, they help to unravel the relevant interaction mechanisms and, on the other hand, allow for the study of sufficiently large systems.  A prominent example is the squirmer model introduced by Lighthill \cite{ligh:52} and revised by Blake \cite{blak:71}. Originally, it was intended as a model for ciliated microswimmers, such as {\em Paramecia}. Nowadays, it is considered as a generic model for a broad class of microswimmers, ranging from  diffusiophoretic particles \cite{hows:07,erbe:08,volp:11} to biological cells ({\em E. coli, Chlamydomonas}, etc.) and has been applied to study collective effects in bulk \cite{llop:10,goet:10a,ishi:06,ishi:07,evan:11,alar:13,moli:13}, at surfaces \cite{ishi:08,llop:10,ishi:13},  and in thin films \cite{zoet:14}.

In its simplest form, a squirmer is represented as a  spherical rigid colloid with a prescribed surface velocity \cite{ligh:52,blak:71,ishi:06}.
Restricting the surface velocity to be tangential, the spherical squirmer is typically characterized by two modes accounting for its swimming velocity and its force-dipole. The latter distinguishes between pushers, pullers, and neutral squirmers. The assumption of a spherical shape is adequate for swimmers like {\em Volvox}, however, the shape of bacteria such as {\em E. coli} or the time-averaged shape of cells such as {\em Chlamydomonas} is nonspherical. Hence, an extension of the squirmer concept to spheroidal objects is desirable.  In 1977, Keller and Wu proposed a generalization of the squirmer model to a prolate-spheroidal shape, which resembles real biological microswimmers such as
\textit{Tetrahymenapyriformis}, \textit{Spirostomum ambiguum}, and \textit{Paramecium multimicronucleatum} \cite{kell:77}.
However, that squirmer model accounts for the swimming mode only and does not include a force-dipole mode. This is unfortunate, since the force-dipole mode determines swimmer-swimmer and swimmer-wall interactions \cite{ishi:07,goet:10a,berk:08,li:14}. A route to incorporate the force-dipole mode into the spheroidal squirmer model was proposed in Ref.~\onlinecite{ishi:13}. However, to the best of our knowledge, the resulting hydrodynamic model is not solvable analytically. In this article, we propose an alternative model for a spheroidal squirmer, taking into account both, a swimming and a force-dipole mode. The major advantage of our approach is that the flow field can be determined analytically (cf. Fig. \ref{fig:theo_SF_BF}).

Various mesoscale simulation techniques have been applied to study the dynamics of squirmers embedded in a fluid, comprising Stokesian dynamics \cite{ishi:07,ishi:08,evan:11}, the boundary-element method  \cite{ishi:06,ishi:13,kyoy:15,li:14,spag:12}, the  multiparticle collision dynamics (MPC) approach \cite{goet:10a,zoet:12,zoet:14}, lattice Boltzmann  simulations \cite{alar:13,pago:13}, the smoothed profile method \cite{moli:13}, and the force-coupling approach \cite{delm:15}. In the following, we will apply the MPC method. MPC is a particle-based simulation technique which incorporates thermal fluctuations \cite{male:99,kapr:08,gomp:09}, provides hydrodynamic correlations \cite{tuez:06,huan:12}, and is easily coupled with other simulation techniques such as molecular dynamics simulations for embedded particles \cite{kapr:08,gomp:09}. The method has successfully been applied in various studies of active systems underlining the importance of hydrodynamic interactions for microswimmers \cite{elge:15,reig:12,reig:13,ruec:07,yang:11,kapr:08,elge:09,earl:07,elge:10,elge:13,thee:13,zoet:14,hu:15,hu:15.1,goet:10a}.

Here, we implement our spheroidal squirmer model in MPC. More specifically, we study the resulting flow field and compare it with the theoretical prediction. Moreover, we present results for the cooperative swimming behavior of two spheroidal squirmers in thin films. Two pullers exhibit a long-time stable configuration, where they swim together in a wedge-like conformation with a constant small angle due to the hydrodynamic interaction between the anisotropic squirmers as well as squirmers and walls. The cooperative and collective swimming motion of spheroidal squirmers in Stokes flow has been addressed in Ref.~\onlinecite{kyoy:15} by an adopted boundary-element method. This approach neglects thermal fluctuations and tumbling of the squirmers completely; only hydrodynamic and excluded-volume interactions determine the squirmer motion. In contrast, our simulation approach includes thermal fluctuations, which affects the stability of the cooperative swimming motion due to the rotational diffusion of a spheroid.

\section{Hydrodynamic model of a spheroidal squirmer} \label{sec:HydrodynamicModelSpheroidalSquirmer}

\subsection{Spheroid geometry}

\begin{figure}
\includegraphics*[width=\columnwidth]{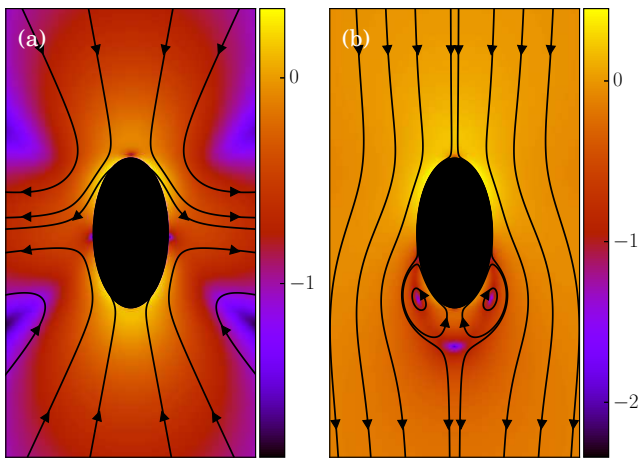}
\caption{\label{fig:theo_SF_BF}
  Flow field of a spheroidal puller with $\beta=3$, (a) in the laboratory frame, and (b) in the body-fixed frame. The logarithm of the magnitude of the velocity field is color coded.
}
\end{figure}

\begin{figure}
\begin{center}
\includegraphics*[width=\columnwidth]{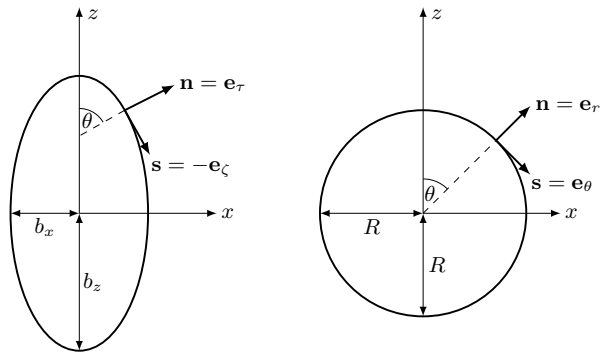}
\caption{\label{fig:sketch_spheroid}
  Sketch of normal and tangent vectors of a spheroidal (left) and spherical (right) squirmer. In the squirmer model, self-propulsion (in $z$-direction) is achieved by a prescribed tangential surface velocity in direction of the tangent vector $\bm{s}$.
}
\end{center}
\end{figure}

We describe a nonspherical squirmer as a prolate spheroidal rigid body with a prescribed surface velocity $\bm{u}_{sq}$. In Cartesian coordinates  $(x,y,z)$, the surface equation of a spheroid, or ellipsoid of revolution, is
\begin{align} \label{Eq:spheroid_surface}
  (x^2+y^2)/b_x^2+z^2/b_z^2=1,
\end{align}
with $b_z$ and $b_x$ the semi-major and semi-minor axis, respectively, and $b_z \geq b_x$ (cf. Fig. \ref{fig:sketch_spheroid}).
We denote half of the focal length by $c=\sqrt{b_z^2-b_x^2}$, which yields the eccentricity $e=c/b_z$.
Furthermore, we define a swimmer diameter as $\sigma=2b_z$.
In terms of prolate ($b_z>b_x$) spheroidal coordinates $(\zeta, \tau, \varphi)$, the Cartesian coordinates are given by
\begin{align} \label{Eq:xyz_of_tau_zeta_phi} \nonumber
  x &= c \sqrt{\tau^2-1}\sqrt{1-\zeta^2} \cos \varphi , \\
  y &= c \sqrt{\tau^2-1}\sqrt{1-\zeta^2} \sin \varphi , \\ \nonumber
  z&=c \tau \zeta ,
\end{align}
where $-1 \leq \zeta \leq 1$, $1 \leq \tau < \infty$, and $0 \leq \varphi \leq 2 \pi$.
All points with $\tau=\tau_0 \equiv e^{-1}$ lie on the spheroid's surface.
The intersection of the spheroid and a meridian plane, where $\varphi$ is constant, is an ellipse.
The normal $\bm{n}$ and tangent $\bm{s}$ to this ellipse are given by the unit vectors
$\bm{e}_\tau$ and $-\bm{e}_\zeta$, respectively, which follow by partial derivative of Eqs.~(\ref{Eq:xyz_of_tau_zeta_phi}) with respect to the coordinates $\zeta$ and $\tau$.
For $b_x=b_z$, the spheroid becomes a sphere. The spherical coordinates
\begin{align}
   (x,y,z)^T=r({\sin \theta \cos \varphi, \sin \theta \sin \varphi, \cos \theta})^T
\end{align}
are obtained from Eq. (\ref{Eq:xyz_of_tau_zeta_phi})
for $\tau \to \infty, \ c\tau = r$, and $\zeta = \cos \theta$. In this limit, the unit vectors turn into $\bm{e}_\tau \to \bm{e}_r$ and $ \bm{e}_\zeta \to - \bm{e}_\theta$ (cf. Fig. \ref{fig:sketch_spheroid}).
The Lam\'{e} metric coefficients for prolate spheroidal coordinates are
$h_\zeta=c (\tau^2-\zeta^2)^{\frac{1}{2}} (1-\zeta^2)^{-\frac{1}{2}}$,
$h_\tau=c (\tau^2-\zeta^2)^{\frac{1}{2}} (\tau^2-1)^{-\frac{1}{2}}$ and
$h_\varphi=c (\tau^2-1)^{\frac{1}{2}} (1-\zeta^2)^{\frac{1}{2}} $.\\

\subsection{Flow field}

The squirmer is immersed in an incompressible low-Reynolds-number fluid, which is described by the incompressible Stokes equations
\begin{align}
  \eta \Delta \bm{v}-\bm{\nabla}p=0, \hspace{2 ex} \bm{\nabla} \cdot \bm{v}=0.
\end{align}
Here, $\bm{v}(\bm{r})$ is the fluid velocity field, $p(\bm{r})$ the pressure field at the position $\bm r$,  and $\eta$ the viscosity.
In an axisymmetric flow, the velocity field can be expressed by the stream function $\Psi$ as \cite{happ:83}
\begin{align} \label{Eq:v_equals_curl_Psi}
  \bm{v}(\zeta, \tau, \varphi)=\textrm{curl} \left( \frac{1}{h_\varphi} \Psi(\tau, \zeta) \bm{e}_\varphi\right).
\end{align}
The stream function itself satisfies the equation \cite{happ:83}
\begin{align} \label{Eq:Stokes_Stream_Function_Equation}
  E^4 \Psi=0,
\end{align}
with the operator \cite{dass:08}
\begin{align}
  E^2=\frac{1}{c^2(\tau^2-\zeta^2)} \left((\tau^2-1)\frac{\partial^2}{\partial \tau^2}+(1-\zeta^2)\frac{\partial^2}{\partial \zeta^2} \right).
\end{align}
Each function $\Psi$ in the kernel of $E^2$ can be represented as \cite{dass:08}
\begin{align}
  \Psi(\tau,\zeta)= \sum_{n=0}^\infty \sum_{i=1}^4 c_n^i \Theta_n^i(\tau, \zeta ),
\end{align}
with constants $c_n^i$ and the functions
\begin{align}
  \Theta_n^1(\tau, \zeta) &= G_n(\tau) G_n(\zeta) ,
  \quad \Theta_n^2(\tau, \zeta) = G_n(\tau) H_n(\zeta), \nonumber \\
  \Theta_n^3(\tau, \zeta) &= H_n(\tau) G_n(\zeta),
  \quad \Theta_n^4(\tau, \zeta) = H_n(\tau) H_n(\zeta). \nonumber
\end{align}
Here, $G_n(x)$ and $H_n(x)$ are Gegenbauer functions of the first and second kind, respectively (see Appendix \ref{app:Gegenbauer}).
The velocity components follow from the stream function via \cite{happ:83}
\begin{align}
  v_\tau&= \frac{1}{h_\zeta h_\varphi} \frac{\partial \Psi}{\partial \zeta}=c^{-2}(\tau^2-1)^{-\frac{1}{2}} (\tau^2-\zeta^2)^{-\frac{1}{2}} \frac{\partial \Psi}{\partial \zeta}  \label{Eq:v_tau_of_stream_function}, \\
  v_\zeta&=-\frac{1}{h_\tau h_\varphi} \frac{\partial \Psi}{\partial \tau}=-c^{-2}(1-\zeta^2)^{-\frac{1}{2}} (\tau^2-\zeta^2)^{-\frac{1}{2}} \frac{\partial \Psi}{\partial \tau} \label{Eq:v_zeta_of_stream_function}.
\end{align}

An important feature of a squirmer is the hydrodynamic boundary condition at its surface, which demands $\bm{v}(\bm{r})=\bm{u}_{sq}$.
For the squirming velocity $\bm{u}_{sq}$ we propose
\begin{align}
  \bm{u}_{sq}&=-B_1 (\bm{s} \cdot \bm{e}_z) \bm{s} -B_2 \zeta (\bm{s} \cdot \bm{e}_z)   \bm{s} \label{Eq:Def_usq_of_zeta}\\
  &=-B_1 \left(1+\beta \zeta \right) (\bm{s} \cdot \bm{e}_z) \bm{s} \label{Eq:Squirming_vel_spheroid_analytical_new} \\
  &=-B_1 \tau_0 (1-\zeta^2)^{\frac{1}{2}}(\tau_0^2-\zeta^2)^{-\frac{1}{2}} \left( 1+\beta \zeta \right)\bm{e}_\zeta.
\end{align}
Here, $\bm{s}$ is the tangent vector, $\bm{e}_z=(0,0,1)^T$ is the unit vector in $z$-direction, $B_1$ and $B_2$ are the two surface velocity modes,  and $\beta=B_2/B_1$ (cf. Fig. \ref{fig:sketch_spheroid}).
$B_1$ determines the swimming velocity, while the $B_2$ term introduces a force-dipole, or pusher ($B_2<0$) and puller ($B_2>0$) mode.
Note that the spherical squirmer introduced by Lighthill and Blake with modes $B_1$ and $B_2$ \cite{ligh:52,blak:71} is recovered for the spherical limit of a spheroid, where
$\zeta \to \cos(\theta)=\bm{n} \cdot \bm{e}_z$. \\

For $B_2=0$, this model of a spheroidal squirmer was already introduced and analysed in Refs.~\onlinecite{kell:77} and \onlinecite{lesh:07}.
An additional force-dipole mode has been introduced in Refs.~\onlinecite{ishi:13} and \onlinecite{kyoy:15} as $\bm{u}_{sq}(\zeta)=-B_1 \bm{s} \cdot \bm{e}_z \left(1+\beta \bm{n} \cdot \bm{e}_z \right) \bm{s}$.
However, we prefer the squirming velocity introduced in Eq.~(\ref{Eq:Squirming_vel_spheroid_analytical_new}), since it yields an analytically solvable boundary value problem
for the Stokes equation.
The two approaches provide a somewhat different flow field in the vicinity of the squirmer,
but both yield the model of Lighthill and Blake in the limit of zero eccentricity.

In the swimmer's rest frame, and with Eq.~(\ref{Eq:Squirming_vel_spheroid_analytical_new}), the boundary value problem becomes
\begin{align}
  \Psi(\tau,\zeta) &\to \frac{1}{2} U_0 c^2 (\tau^2-1)(1- \zeta^2)  \textrm{ for } \tau \to \infty,  \label{Eq:Spheroid_Squirmer_BC1}  \\
  \Psi(\tau_0,\zeta) &= 0  \textrm{ for all } \zeta \label{Eq:Spheroid_Squirmer_BC2}, \\
   \left. \frac{\partial \Psi}{\partial \tau}\right|_{\tau=\tau_0}  &=
   (B_1 +B_2 \zeta)c^2 \tau_0 (1-\zeta^2) \textrm{ for all } \zeta. \label{Eq:Spheroid_Squirmer_BC3}
\end{align}
Equation (\ref{Eq:Spheroid_Squirmer_BC1}) implies a constant background flow $\bm{v}=-U_0\bm{e}_z$ infinitely far from the squirmer, Eq.~(\ref{Eq:Spheroid_Squirmer_BC2}) guarantees $v_\tau=0$ at the spheroid surface, and Eq.~(\ref{Eq:Spheroid_Squirmer_BC3}) demands $v_\zeta=\bm{u}_{sq}(\zeta) \cdot \bm{e}_\zeta$.
Due to linearity of the Stokes stream function equation (\ref{Eq:Stokes_Stream_Function_Equation}), we can solve this boundary value problem for $B_2=0$ first, which yields the stream function $\Psi_1$. Subsequently we solve the problem
\begin{align}
  \Psi(\tau,\zeta)&   \textrm{ converges for } \tau \to \infty,  \label{Eq:Spheroid_Squirmer_B2_BC1}  \\
  \Psi(\tau_0,\zeta) &= 0 \textrm{ for all } \zeta \label{Eq:Spheroid_Squirmer_B2_BC2}, \\
    \left. \frac{\partial \Psi}{\partial \tau}\right|_{\tau=\tau_0} &=
    -c^2(\tau_0^2-\zeta^2)^{\frac{1}{2}} (1-\zeta^2)^{\frac{1}{2}} u_2(\zeta), \nonumber \\
    &=B_2 c^2 \tau_0 (1-\zeta^2) \zeta   \textrm{ for all } \zeta. \label{Eq:Spheroid_Squirmer_B2_BC3}
\end{align}
Equation (\ref{Eq:Spheroid_Squirmer_B2_BC1}) imposes a vanishing velocity field infinitely far from the squirmer, Eq. (\ref{Eq:Spheroid_Squirmer_B2_BC2}) again guarantees $v_\tau=0$
at the spheroid surface, and Eq. (\ref{Eq:Spheroid_Squirmer_B2_BC3}) demands $v_\zeta=\bm{u}_{sq}(\zeta,B_1=0) \cdot \bm{e}_\zeta$.
We denote the solution of the problem Eqs.~(\ref{Eq:Spheroid_Squirmer_B2_BC1})-(\ref{Eq:Spheroid_Squirmer_B2_BC3}) by $\Psi_2$. Finally, $\Psi=\Psi_1+\Psi_2$ solves the
initial problem (\ref{Eq:Spheroid_Squirmer_BC1})-(\ref{Eq:Spheroid_Squirmer_BC3}) for arbitrary $B_1$ and $B_2$.

The boundary value problem Eqs.~(\ref{Eq:Spheroid_Squirmer_BC1})-(\ref{Eq:Spheroid_Squirmer_BC3}) for $B_2=0$ can be solved by the ansatz
\begin{align} \label{Eq:psi1_for_neutral_spheroidal_squirmer}
  \Psi_1(\tau,\zeta)=\alpha_1 G_2(\tau) G_2(\zeta)+\alpha_2 H_2(\tau)G _2(\zeta)+\alpha_3 \tau (1-\zeta^2).
\end{align}
Here, the third term is found by the separation ansatz $\Psi(\tau,\zeta)=g(\tau)(1-\zeta^2)$ for Eq. (\ref{Eq:Stokes_Stream_Function_Equation}).
Equation (\ref{Eq:Spheroid_Squirmer_BC1}) directly yields $\alpha_1=-2 U_0 c^2$. The remaining coefficients $\alpha_2$ and $\alpha_3$ are determined by Eqs.~(\ref{Eq:Spheroid_Squirmer_BC2}) and (\ref{Eq:Spheroid_Squirmer_BC3}), keeping in mind that $B_2=0$. This yields
\begin{align} \label{eq:alpha2}
  \alpha_2&=2c^2 \frac{ U_0(\tau_0^2+1)-2B_1 \tau_0^2}{(\tau_0^2+1) \coth^{-1} \tau_0 -\tau_0}, \\[2 ex] \label{eq:alpha3}
  \alpha_3&=c^2 \frac{B_1 \tau_0(\tau_0-(\tau_0^2-1)\coth^{-1} \tau_0 ) -U_0  }{(\tau_0^2+1) \coth^{-1} \tau_0 -\tau_0}.
\end{align}
The boundary value problem Eqs.~(\ref{Eq:Spheroid_Squirmer_B2_BC1})-(\ref{Eq:Spheroid_Squirmer_B2_BC3}) can be solved by the ansatz
\begin{align}
  \Psi_2(\tau,\zeta)=\alpha_4 G_3(\tau)G_3(\zeta) + \alpha_5 H_3(\tau) G_3(\zeta)+\alpha_6 \zeta (1-\zeta^2).
\end{align}
As before, the third term follows by a separation ansatz $\Psi(\tau,\zeta)=g(\tau) \zeta(1-\zeta^2)$ for Eq. (\ref{Eq:Stokes_Stream_Function_Equation}).
Equation~(\ref{Eq:Spheroid_Squirmer_B2_BC1}) yields $\alpha_4 =0$. The coefficients $\alpha_5$ and $\alpha_6$ are
determined by Eqs. (\ref{Eq:Spheroid_Squirmer_B2_BC2})-(\ref{Eq:Spheroid_Squirmer_B2_BC3}) such that
\begin{align}
  \alpha_5&=c^2\frac{4B_2 \tau_0}{3 \tau_0+(1-3 \tau_0^2)\coth^{-1} \tau_0} \label{Eq:alpha_5_equals}, \\[2 ex]
  \alpha_6&=c^2 B_2 \tau_0 \frac{ 2/3-\tau_0^2+\tau_0(\tau_0^2-1)\coth^{-1} \tau_0 }{3 \tau_0+(1-3 \tau_0^2)\coth^{-1} \tau_0}. \label{Eq:alpha_6_equals}
\end{align}
The total stream function $\Psi=\Psi_1+\Psi_2$ can be transformed to the laboratory frame (cf. Fig. \ref{fig:theo_SF_BF}) by adding the background flow $\bm{v}=U_0\bm{e}_z$, which yields
\begin{align} \label{Eq:Stream_Function_Lab_Frame}
  \Psi^{lab}&=\Psi-\frac{1}{2} U_0 c^2(\tau^2-1)(1-\zeta^2) \\
  &=\alpha_2 H_2(\tau)G _2(\zeta)+\alpha_3 \tau (1-\zeta^2) \nonumber \\
  &\quad + \alpha_5 H_3(\tau) G_3(\zeta)+\alpha_6 \zeta (1-\zeta^2).
\end{align}
The force by the fluid on the spheroid is given by \cite{happ:83}
\begin{align}
  F_z=\lim_{r \to \infty} \frac{r \Psi^{lab}}{\bar{r}^2} = 8\pi \eta \alpha_3 /c,
\end{align}
where $r=\sqrt{x^2+y^2+z^2}$ and $\bar{r}=\sqrt{x^2+y^2}$. As expected, $\Psi_2$ does not contribute to the
force, since it assumes a constant value at infinity.
Since a swimmer must be force free,  $F_z=0$, which implies $\alpha_3=0$. Then, Eq.~(\ref{eq:alpha3}) yields  the swimming velocity of the squirmer ($\tau_0=1/e$)
\begin{align} \label{Eq:Swimming_velocity_spheroid_squirmer}
  U_0=B_1 \tau_0(\tau_0-(\tau_0^2-1)\coth^{-1} \tau_0 ),
\end{align}
which was already found by Keller and Wu for the case $B_2=0$ \cite{kell:77}.
As a consequence, $\alpha_2$ in Eq. (\ref{eq:alpha2}) simply becomes $\alpha_2=2B_1 c^2 \tau_0(\tau_0^2-1)$.\\

The flow field of a point-like force-dipole is given by \cite{spag:12,elge:15}
\begin{align}
  \bm{v}^{FD}=\frac{P}{8 \pi \eta} \frac{\bm{r}}{r^3} \left(\frac{3z^2}{r^2} -1 \right),
\end{align}
with the dipole strength $P$,  whereas the flow field of a source doublet is \cite{spag:12}
\begin{align}
  \bm{v}^{SD}=\kappa \frac{1}{r^3} \left(-\bm{e}_z+\frac{3z \bm{r}}{r^2} \right),
\end{align}
with the source-doublet strength $\kappa$.
Comparing the corresponding stream functions with Eq. (\ref{Eq:Stream_Function_Lab_Frame}) far from the origin, we find
\begin{align}
  P&=-8 \pi \eta \alpha_6, \\
  \kappa&=-\frac{c \alpha_2}{6}=-\frac{B_1}{3}c^3 \tau_0(\tau_0^2-1)
\end{align}
for our model. As expected, in the spherical limit ($b_z \to b_x \equiv R$, where $R$ is the radius) we obtain $P=-4\pi \eta B_2 R^2$ and $\kappa=-B_1R^3/3$.

\begin{figure}
\includegraphics*[width=\columnwidth]{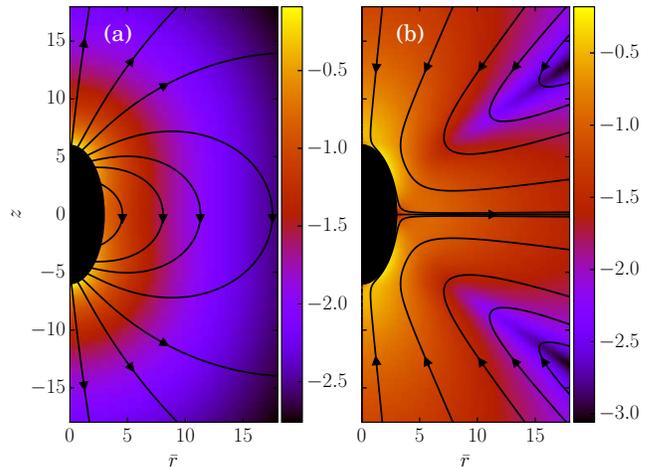}
\caption{\label{fig:theo_B1_B2}
Fluid velocity fields of a spheroidal squirmer in the laboratory frame for (a) $B_1=1, \ B_2=0$,  and (b) $B_1=0, \ B_2=1$.
The corresponding stream function is given by Eq. (\ref{Eq:Stream_Function_Lab_Frame}).
The logarithm of the magnitude of the velocity field is color coded. Note that the pusher velocity field with $B_1=0, \ B_2=-1$ is not shown, since it follows from that of the puller with $B1=0, \ B_2=1$ by inverting the arrows.
}
\end{figure}

Examples of fluid velocity fields of a spheroidal squirmer are presented  in Figs.~\ref{fig:theo_SF_BF} and \ref{fig:theo_B1_B2}.

\section{Multiparticle collision dynamics} \label{sec:MPC}

Multiparticle collision dynamics (MPC) is a stochastic, particle based mesoscale hydrodynamic simulation method \cite{gomp:09}.
Thereby, a fluid is modeled by $N$ point particles with equal mass $m$, undergoing subsequent streaming and collision steps.
In the streaming step, the particle positions $\bm{r}_i, \ i=1,\dots,N$, are updated according to
\begin{align}
  \bm{r}_i(t+h)=\bm{r}_i(t)+h\bm{v}_i(t),
\end{align}
where $\bm{v}_i$ are the particle velocities and $h$ is denoted as collision time step.
In the subsequent collision step, the particle velocities are changed by a stochastic process, which mimics internal fluid interactions. In order to define the local collision environment,
particles are sorted into cells of a cubic lattice with lattice constant $a$.
Different realizations for this stochastic process have been proposed.\cite{male:99,alla:02,nogu:07}
We employ the stochastic rotation dynamics (SRD) approach of MPC with angular momentum conservation (SRD+a) \cite{nogu:08,thee:15}, which updates the particle velocities in a cell according to
\begin{align} \label{Eq:MPC_Collision_Step}
\begin{split}
 {\bm v}_i^{new} = & ~{\bm v}_{cm}+{\bm R}(\alpha){\bm v}_{i,c} - {\bm r}_{i,c} \times \\
	  & \times \Big[ m {\bm I}^{-1} \sum_{j \in cell}\left\{{\bm r}_{j,c}\times \left({\bm v}_{j,c}-{\bm R}(\alpha) {\bm v}_{j,c} \right) \right\} \Big] .
\end{split}
\end{align}
Here, ${\bm r}_{i,c}={\bm r}_{i}-{\bm r}_{cm}$, where ${\bm r}_{cm}$ is the center-of-mass position of the particles in the cell, and similarly, ${\bm v}_{i,c}={\bm v}_{i}-{\bm v}_{cm}$, with the center-of-mass velocity ${\bm v}_{cm}$.
$\bm{R}(\alpha)$ is the rotation matrix, which describes a rotation around a randomly oriented axis by the angle $\alpha$. The angle $\alpha$ is a constant, and the axis of rotation is chosen independently for each cell and time step. Finally, ${\bm I}$ is the moment-of-inertia tensor of the particles in the center-of-mass reference frame of the cell. Partition of the system into collision cells leads to a violation of Galilean invariance. To reestablish Galilean invariance, a random shift of the collision-cell lattice is introduced at every collision step \cite{ihle:01,ihle:03}.

Since energy is not conserved in the collision step, we apply a cell level canonical thermostat at temperature $T$ \cite{huan:10,huan:15}. The latter ensures Maxwell-Boltzmann distributed velocities.
The MPC algorithm is embarrassingly parallel. Hence, we implement it on a Graphics Processing Unit (GPU) for a high performance gain \cite{west:14}.

The following simulations are performed with the mean number of particles per collision cell $\left< N_c \right>=10$, the rotation angle $\alpha=130^\circ$,
and the time step $h=0.02 \sqrt{ma^2/(k_B T)}$, which yields a fluid viscosity of $\eta=17.8 \sqrt{m k_B T/a^4}$.

\section{Implementation of a spheroidal squirmer in MPC}

A spheroidal squirmer is a homogeneous rigid body characterized by its mass $M$, center-of-mass position $\bm{C}$, orientation $\bm{q}$, translational velocity $\bm{U}$, and angular momentum $\bm{l}$.
Thereby, $\bm{q}=(q_0,q_1,q_2,q_3)$ is a rotation quaternion and can be related to the rotation matrix $\bm{D}$, which transforms vectors from the laboratory frame to the body-fixed frame \cite{alle:87}.
We distinguish vectors in the laboratory frame and body-fixed frame by a superscript, i.e., $\bm{v}^s$ is a vector in the laboratory (or space-fixed) frame while
\begin{align} \label{Eq:vb_equals_Dvs}
\bm{v}^b=\bm{D}\bm{v}^s
\end{align}
is the corresponding vector in the body-fixed frame.
For vectors in the laboratory frame, we will frequently omit the superscript.
The orientation vector of a spheroid is $\bm{e}=\bm{D}^T \bm{e}^b=\bm{D}^T (0,0,1)^T$.
The moment of inertia tensor in the body-fixed frame $\bm{I}^b$ is a constant diagonal matrix with diagonal elements $I_x=(M/5)(b_x^2+b_z^2)=I_y$ and $I_z=(2M/5) b_x^2$.
When needed, the angular velocity is calculated as $\bm{\Omega}^s=\bm{D}^T \left( \bm{I}^b \right)^{-1} \bm{D} \bm{l}^s$.\\
For all simulations we choose a neutrally bouyant spheroid, i.e., $M=\rho  (4 \pi/3) b_z b_x^2$, where $\rho$ is the fluid mass density.

\subsection{Streaming step} \label{sec:Streaming}
During the streaming step, a spheroid will collide with several MPC particles.
Since the total change in (angular) momentum of a spheroid during one streaming step is small, we perform the collisions with MPC particles in a coarse-grained way \cite{padd:06}:

For the streaming step at time $t$, we determine the spheroid's position, velocity, orientation, and angular velocity at times
$t+h/2$ and $t+h$, under the assumption that there is no interaction with MPC particles.
However, steric interactions between spheroids, as well as spheroids and walls are taken into account as described in Sec.~\ref{sec:RBD_for_Spheroids}.

Subsequently, all MPC particles are streamed, i.e., their positions are updated according to $\bm{r}_i(t+h)=\bm{r}_i(t)+h\bm{v}_i(t)$. Thereby, a certain fraction of MPC particles penetrates a spheroid. To detect those particles in an efficient way, possible collision cells intersected by the spheroid are identified first. For this purpose, we select all those cells, which are within a sphere of radius $b_z$ enclosing the spheroid instead of the spheroid itself, which is more efficient, since it avoids rotating candidate cells into the body-fixed frame during selection. A loop over all particles in respective collision cells identifies those particles, which are inside the spheroid and they are labeled with the spheroid index. Then, each particle $i$ inside a spheroid at time $t+h$ is moved back in time by half a time step and subsequently translated onto the spheroid's surface.
The translation can be realized in different ways. One possibility is to constructing a virtual spheroid with semi-axes $\tilde b_z, \ \tilde b_x, \ \tilde b_z/\tilde b_x=b_z/b_x$ and $\bm{r}_i(t+h/2)$ on its surface. The particle is then translated along the normal vector of the virtual spheroid until it is on the real spheroid's surface.
Alternatively, the difference vector $\bm{r}_i(t+h/2)-\bm{C}(t+h/2)$ can be scaled such that the particle position lies on the spheroid'surface. We tried both approaches and found no significant difference.
Once the MPC particle at time $t+h/2$ is located on the spheroid's surface, the momentum transfer
\begin{align} \label{Eq:J_i_equals}
\bm{J}_i=2m\left\{ \bm{v}_i- \bm{U}-\bm{\Omega}\times (\bm{r}_i-\bm{C}) -\bm{D}^T\bm{u}_{sq}^b [\bm{D} (\bm{r}_i - \bm{C}) ] \right\}
\end{align}
at time $t+h/2$ is determined, taking into account the squirmer surface fluid velocity $\bm{u}_{sq}$  of Eq.~(\ref{Eq:Def_usq_of_zeta}) \cite{down:09}.
Thereby, a useful identity to determine $\bm{s}$ is given in Eq. (8) of Ref.~ \onlinecite{kell:77}, and $\zeta$ is given by
\begin{align}
 \zeta =\frac{1}{2c}\left( \sqrt{x^2+y^2+(z+c)^2}-\sqrt{x^2+y^2+(z-c)^2} \right).
\end{align}
The velocity of the MPC particle is updated according to
$\bm{v}_i'=\bm{v}_i-\bm{J}_i/m$. Subsequently, the position $\bm{r}_i(t+h)$ is obtained by streaming the MPC particle for the remaining time $h/2$ with velocity $\bm{v}_i'$, i.e.,
$\bm{r}_i(t+h)=\bm{r}_i(t+h/2)+h\bm{v}_i'/2$.

As a consequence of the elastic collisions, the center-of-mass velocity and rotation frequency of a spheroid are finally given by
\begin{align}
    \bm{U}(t+h)'&=\bm{U}(t+h)+\bm{J}/M, \\
    \bm{\Omega}(t+h)'&=\bm{\Omega}(t+h) +\bm{D}^T \left( \bm{I}^b \right)^{-1} \bm{D} \bm{L} ,
  \end{align}
where $\bm{J}=\sum_i \bm{J}_i$ is total momentum transfer by the MPC fluid  and $\bm{L}=\sum_i \left( \bm{r}_i(t+h/2)-\bm{C}(t+h/2) \right) \times \bm{J}_i$ is the respective
angular momentum transfer.

\subsection{Collision step}

In a first step, ghost particles are distributed inside each spheroid \cite{lamu:01,padd:06}. The number density and mass are equal for ghost and fluid particles.
The ghost particle positions $\bm{r}_i^g$ are uniformly distributed in the spheroid and their velocities are given by
\begin{align}
  \bm{v}_i^g=\bm{U}+\bm{\Omega}\times(\bm{r}_i-\bm{C})+\bm{u}_{sq,i}+\bm{v}_{R,i}.
\end{align}
The Cartesian components of $\bm{v}_{R,i}$ are Gaussian-distributed random numbers with zero mean and variance $\sqrt{k_B T/m}$.
The squirming velocity $\bm{u}_{sq,i}$ is determined by Eq.~(\ref{Eq:Def_usq_of_zeta}), with the  ghost particle position projecting onto the spheroid's surface (cf. Sec.~\ref{sec:Streaming}). 
As a result of MPC collisions, a spheroid's linear and angular momenta change by  $\bm{J}^g_i=m(\bar{\bm v}_i^g-{\bm v}_i^g)$ and
$\bm{L}^g_i=\left( \bm{r}_i^g-\bm{C} \right) \times \bm{J}^g_i$, where $\bar{\bm v}_i^g$ and ${\bm v}_i^{g}$ are the ghost particle's velocity after
and before the MPC collision.
Hence, the spheroid velocity and angular velocity become
\begin{align}
     \bm{U}'&=\bm{U}+\bm{J}^g/M, \\
     \bm{\Omega}'&=\bm{\Omega} +\bm{R}^T \left( \bm{I}^b \right)^{-1} \bm{R} \bm{L}^g.
\end{align}

\subsection{Rigid body dynamics for spheroids} \label{sec:RBD_for_Spheroids}
During the streaming step, the spheroids move according to rigid-body dynamics, governed by \cite{omel:98}
\begin{align}
   M \ddot{\bm{C}}&= \bm{F}, \\
   \ddot{\bm{q}}&=\frac{1}{2} \left[ \bm{Q}(\dot{\bm{q}})  \left( \begin{matrix} 0 \\ \bm{\Omega}^b \end{matrix} \right)
  + \bm{Q}(\bm{q}) \left( \begin{matrix} 0 \\ \dot{\bm{\Omega}}^b \end{matrix} \right) \right], \\
  \dot{\bm{q}}&=\frac{1}{2} \bm{Q}(\bm{q}) \left( \begin{matrix} 0 \\ \bm{\Omega}^b \end{matrix} \right), \label{Eq:qdot_equals} \\
  \frac{d \Omega_\alpha^b}{dt}&=I_\alpha^{-1} \left[ T_\alpha^b+(I_\beta - I_\gamma) \Omega_\beta^b \Omega_\gamma^b \right]. \label{Eq:EulerRotation}
\end{align}
Here, $\bm{Q}(\bm{q})$ is defined in Eq. (\ref{Eq:def_Q_of_q}), and
$\bm{F}$ and $\bm{T}$ are the force and torque acting on the spheroid. Forces and torques are derived from steric interaction potentials as presented in Appendix \ref{app:steric_interactions}.
Equations (\ref{Eq:EulerRotation}) are Euler's equations for rigid body dynamics and hold for $(\alpha, \beta, \gamma)=(x,y,z), (y,z,x)$, and $ (z,x,y)$.
Whenever necessary, body-fixed and laboratory-frame quantities can be related by the rotation matrix $\bm{D}$ which is given in terms of the quaternion $\bm{q}$ in Eq. (\ref{Eq:Rotation_matrix_for_RBD}).

For the numerical integration of the equations of motion, the widely applied
leap-frog method \cite{alle:87} is not useful, since velocity, angular momentum, position, and orientation are required at the same point in time for the coupling to the MPC method.
Hence, we employ the Verlet algorithm for rigid-body rotational motion proposed in Ref.~ \onlinecite{omel:98}.
Integration for a time step $\tau$ is performed as follows:
\begin{itemize}
\item[(i)] Update $\bm{C}$ and $\bm{q}$ according to (cf. Eqs.~(\ref{Eq:qdot_equals}) and (\ref{Eq:EulerRotation}))
\indent \begin{align}
    \bm{C}(t+\tau)&=\bm{C}(t)+\bm{U}(t) \tau + \frac{\tau^2}{2M} \bm{F}^s(t),\\
    \bm{q}(t+\tau)&= (1-\tilde{\lambda}) \bm{q}(t)+\dot{\bm{q}}\tau+\frac{\tau^2}{2} \ddot{\bm{q}},\\
     \tilde{\lambda}&= 1-\dot{\bm{q}}^2 \tau^2/2  \nonumber \\
      & \quad - \sqrt{1-\dot{\bm{q}}^2 \tau^2 - \dot{\bm{q}} \cdot \ddot{\bm{q}}\tau^3-(\ddot{\bm{q}}^2-\dot{\bm{q}}^4) \tau^4/4  }.
\end{align}
\noindent The parameter $\tilde{\lambda}$ is introduced to guarantee $\bm{q}^2=1$.
\item[(ii)] Calculate forces and torques $\bm{F}^s(t+\tau)$ and $\bm{T}^s(t+\tau)$.
\item[(iii)] Update $\bm{U}$ and $\bm{l}^s$ according to
\begin{align}
    \bm{U}(t+\tau) & =\bm{U}(t)+\frac{\tau}{2M}[\bm{F}^s(t)+\bm{F}^s(t+\tau)],\\
    \bm{l}^s(t+\tau) & =\bm{l}^s(t)+\frac{\tau}{2}[\bm{T}^s(t)+\bm{T}^s(t+\tau)].
\end{align}
\end{itemize}

\section{Simulations -- thermal properties and flow field}
\subsection{Passive colloid}
%
\begin{figure}
\includegraphics*[width=\columnwidth]{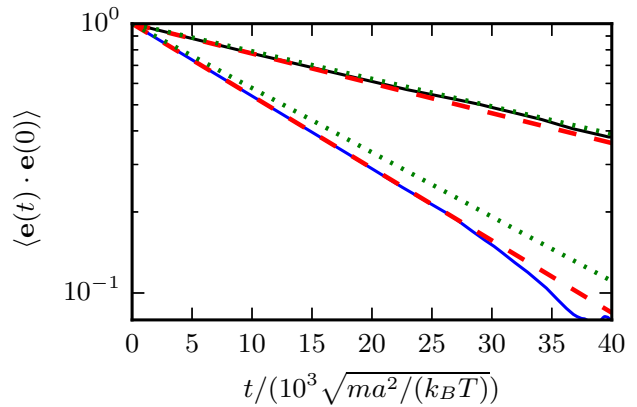}
\caption{\label{fig:orientation_correlation}
  Orientation correlation functions $\langle \bm{e}(t) \cdot \bm{e}(0) \rangle$ for passive spheroids with $b_z=6a,b_x=3a$ (bottom blue line)
  and $b_z=9a,b_x=3a$ (top black line).
  The plot shows the simulation data (blue and black solid lines), an exponential fit to that data (red dashed),
  and the theoretical prediction according to Eq.~(\ref{Eq:Exp_Decay_of_orientational_correlation_passive}) (green dotted).
}
\end{figure}
For the passive spheroidal colloid ($B_1=B_2=0$), we perform equilibrium simulations and determine $\langle U_\alpha^2 \rangle$ as well as $\langle (\Omega^b_\alpha)^2 \rangle$ for $\alpha \in \{x,y,z \}$.
Due to the equipartition of energy, we expect
\begin{align}
  \langle U_\alpha^2 \rangle&=\frac{k_B T}{M}, \label{Eq:equi_u}\\
  \langle (\Omega^b_\alpha)^2 \rangle&=\frac{k_B T}{I_\alpha}. \label{Eq:equi_omega}
\end{align}
We fix the aspect ratio $b_z/b_x=2$ and vary $b_x$ in the range $b_x \in [2a, 4a]$. The simulation results agree very well with the theoretical values (\ref{Eq:equi_u}) and (\ref{Eq:equi_omega}). As expected, the deviations from theory decrease with increasing spheroid size, due to a better resolution in terms of collision cells.
In general, the relative error $\sigma_r=(\langle x_{theo}^2\rangle -\langle x_{sim}^2\rangle)/\langle x_{theo}^2\rangle$ is larger for $\Omega^b_\alpha$ than for $U_\alpha$.
We find the largest relative error for $\langle (\Omega^b_z)^2 \rangle$, namely $\sigma_r = 9.5\%, 5.3\%,$ and $3.1\%$ for $b_x=2a,3a,$ and $4a$. Hence, we choose the minor axis $b_x\geq 3a$ in the following.
%
%

In addition, we determine the orientation correlation function $\langle \bm{e}(t) \cdot \bm{e}(0) \rangle$.
The theory of rotational Brownian motion \cite{favr:60} predicts
\begin{align} \label{Eq:Exp_Decay_of_orientational_correlation_passive}
  \langle \bm{e}(t) \cdot \bm{e}(0) \rangle=\exp \left( -2 D_R^\perp t \right),
\end{align}
where $D_R=(2 D_R^\perp +D_R^\parallel)/3$, $D_R^\parallel=k_B T/\xi^\parallel$, $D_R^\perp=k_B T/\xi^\perp$, and $\xi^\parallel$ and $\xi^\perp$ are
the parallel and perpendicular rotational friction coefficients of a prolate spheroid with respect to the major semi-axis; explicitly \cite{kim:13}
\begin{align}
  \xi^\parallel&=8 \pi \eta b_z^3  \frac{4}{3}e^3(1-e^2)(2e-(1-e^2)L)^{-1}, \\
  \xi^\perp&=8 \pi \eta b_z^3  \frac{4}{3}e^3(2-e^2)(-2e+(1+e^2)L)^{-1}, \\
  L&=\log \left( \frac{1+e}{1-e} \right)
\end{align}
Simulation results for the orientational auto-correlation function are shown in Fig. \ref{fig:orientation_correlation} for two spheroids of different eccentricity. The correlation functions decay exponentially. However, for the spheroid with the smaller eccentricity, we find a somewhat faster decay than predicted by theory, whereas good agreement is found for the larger spheroid. We attribute the difference to finite-size effects related to the discreteness of the collision lattice. For larger objects, discretization effects become smaller.

\subsection{Squirmer}
\begin{figure}
\includegraphics*[width=\columnwidth]{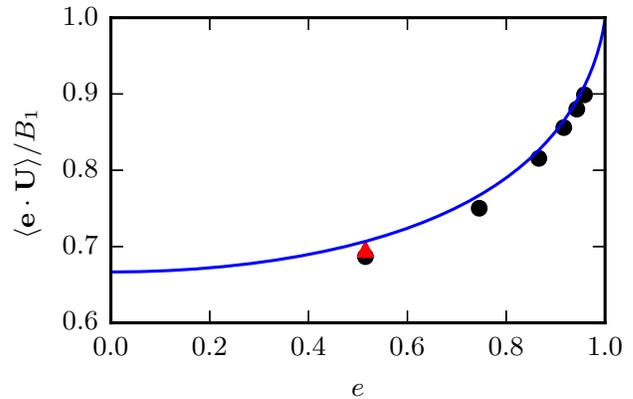}
\caption{\label{fig:U_of_tau0}
  Mean swimming velocity as function of the eccentricity $e$ for a spheroidal squirmer with $B_1=0.05 \sqrt{k_B T/m}$ and $B_2=0$. The solid line shows the theoretical prediction of Eq. (\ref{Eq:Swimming_velocity_spheroid_squirmer}).
  Black dots are simulation results. The eccentricity was varied by changing $b_z$ and keeping $b_x=3a$ constant. For the red triangle,
  we simulated a larger spheroid with $b_x=6a$, which shows a better agreement with theory.
}
\end{figure}

\begin{figure}
\includegraphics*[width=\columnwidth]{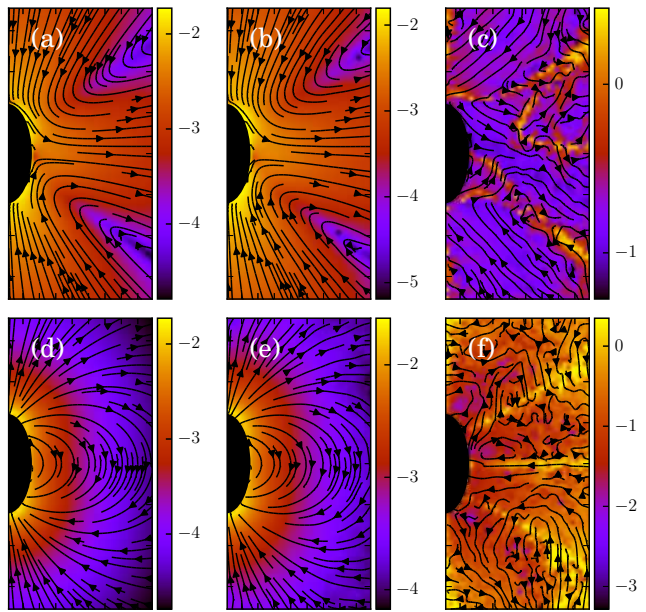}
\caption{\label{fig:squirmer_sim_theo}
   Fluid flow fields of a spheroidal squirmer in the laboratory frame with $b_x=3a$, $b_z=6a$, $B_1=0.01 \sqrt{k_BT/m}$, and $\beta=3$ ((a),(b),(c)), and with $B_1=0.05 \sqrt{k_BT/m},\beta=0$ ((d),(e),(f)). The logarithm of the magnitude of the velocity field (in units of $\sqrt{k_BT/m}$) is color coded.
  The plots  (a), (d) show theoretical results, (b), (e) simulation results,  and (c), (f) relative errors.
  The relative error of the flow field is defined as $\Delta v_\alpha=| v_\alpha^{theo}-v_\alpha^{sim}|/[(|v_\alpha^{theo}|+|v_\alpha^{sim}|)/2]$.
  Note, due to the discrete representation of the velocity field, some streamlines end abruptly.
}
\end{figure}

%
We determine the steady state swimming velocity of a squirmer via $\langle \bm{e} \cdot \bm{U}\rangle$, which should be equal to $U_0$ (cf. Eq. (\ref{Eq:Swimming_velocity_spheroid_squirmer})).
Results for various eccentricities are displayed in Fig.  \ref{fig:U_of_tau0}. The velocity $U_0$ increases with increasing eccentricity $e$ in close agreement with the theoretical prediction of Eq.~(\ref{Eq:Swimming_velocity_spheroid_squirmer}).
We confirm that the force-dipole parameter $\beta$ does not affect the velocity of the squirmer, as long as the Reynolds number $\textrm{Re}$ is low, i.e., $\textrm{Re}=\rho U_0 b_z/\eta  \lesssim 0.1$.
We also determine the orientational correlation function and find that a squirmer exhibits the same orientational decorrelation as the corresponding passive particle (cf. Fig. \ref{fig:orientation_correlation}).
%

Moreover, we calculate the flow field from the simulation data and compare it with the theoretical prediction. As shown in Fig.~\ref{fig:squirmer_sim_theo}, the two fields are in close agreement.
The two-dimensional flow field of the MPC fluid, averaged over the rotation angle $\varphi$, is determined at the vertices of a fine resolution mesh.
The velocities at these vertices include averages over time of an individual realization as well as ensemble averages over various realizations. By the latter, we determine an estimate for the error of the mean velocity. The median (over vertices) of this error is approximately $5\%$ for the parameters of Fig. \ref{fig:squirmer_sim_theo} (b) and $10\%$ for that of Fig. \ref{fig:squirmer_sim_theo} (e).
Note that we choose a smaller swimming mode $B_1$ for the puller (Fig. \ref{fig:squirmer_sim_theo} (b)) than for the neutral squirmer (Fig. \ref{fig:squirmer_sim_theo} (e)). The reason is that the agreement with theory was not satisfactory for the puller with $B_1=0.05\sqrt{k_BT/m}$, which we attribute to nonlinear convective effects.
In Figs. \ref{fig:squirmer_sim_theo} (c) and (f), we observe lines of high relative errors (yellow in the color code).
They appear because theory predicts $v_{\bar{r}}=0$ or $v_z=0$ for these lines, which is difficult to achieve in simulations.
Hence, the overall agreement between simulations and theory is very satisfactory, and the implementation is  very valuable  for the simulation of squirmer-squirmer and squirmer-wall interactions, where the details of the flow field matter.

\section{Cooperative swimming in thin films}
\begin{figure}
\includegraphics*[width=\columnwidth]{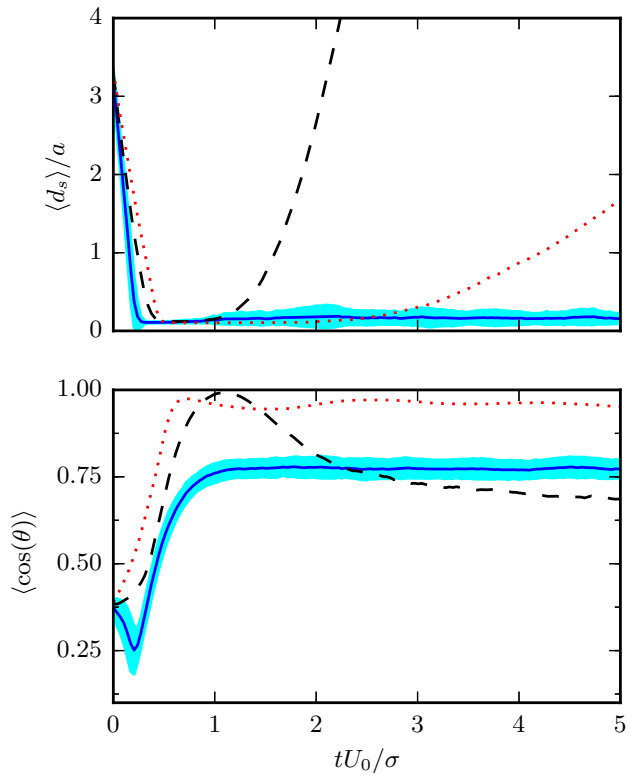}
\caption{\label{fig:d_cos}
  Average surface-to-surface distance $d_s$ and orientation of squirmers, where $\cos(\theta)=\bm{e}_1 \cdot \bm{e}_2$, as function of time.
  The solid blue, dashed black, and dotted red lines correspond to pullers $\beta=4$, neutrals $\beta=0$ and pushers $\beta=-4$.
  The standard deviation of the blue line ($\beta=4$) is indicated by the cyan shaded region.
}
\end{figure}
\begin{figure}
\begin{center}
\hspace{-0.9 ex} \includegraphics*[width=0.885 \columnwidth]{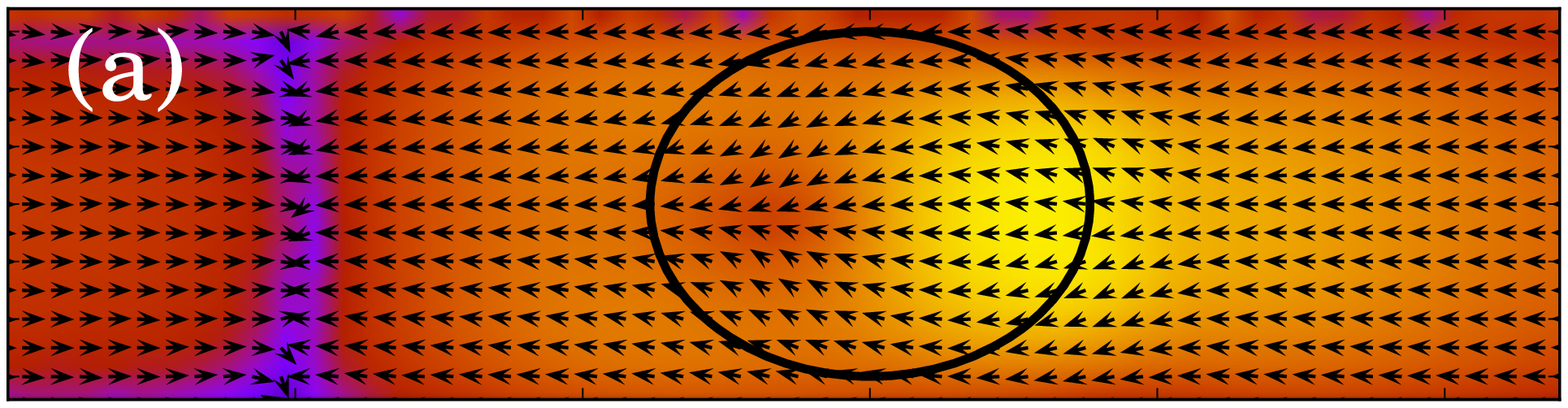}
\includegraphics*[width=0.9 \columnwidth]{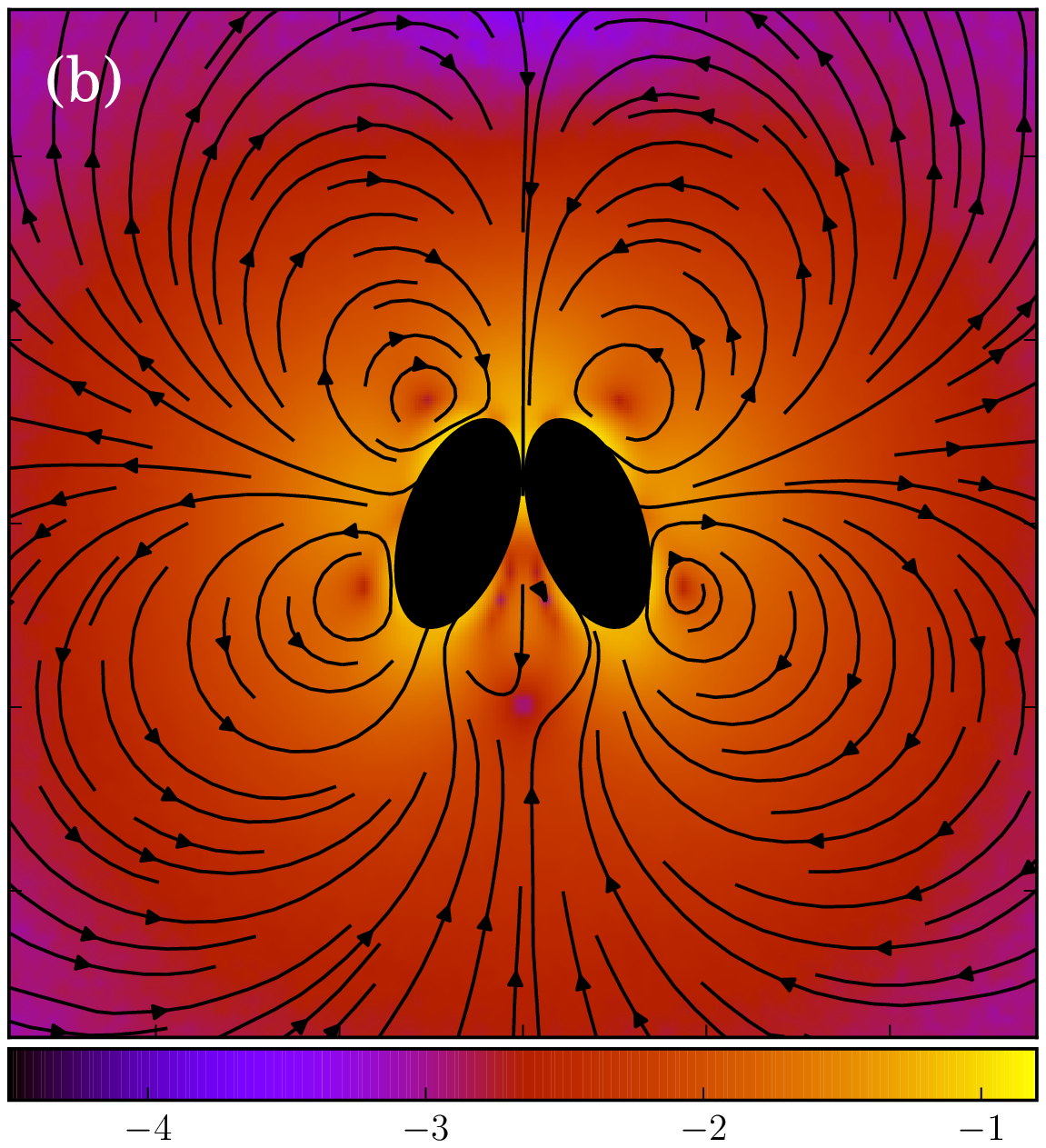}
\caption{\label{fig:coop_flow_field}
  Flow field of two cooperatively swimming pullers in the laboratory frame. The logarithm of the magnitude of the velocity field (in units of $\sqrt{k_BT/m}$) is color coded. 
  We denote the direction normal to the walls by $y$, the cooperative swimming direction by $z$, the remaining Cartesian axis by $x$, and choose the swimmers' center of mass as origin. (a) shows the flow field at $x=0$ in the $zy$-plane, while (b) shows the flow field at $y=0$ in the $xz$-plane. The black elliptical shapes indicate the projection of the swimmers onto the considered plane.
  Note that periodic boundary conditions are employed in the MPC simulation and affect the flow field, which leads to closed flow lines on the length scale of the periodic system.
}
\end{center}
\end{figure}
We simulate the cooperative swimming behavior of two squirmers in a slit geometry. The slit is formed by two parallel no-slip walls located at $y=0$ and $y=L_y$. The no-slip boundary condition is implemented by applying the bounce-back rule and ghost particles of zero mean velocity in the walls \cite{lamu:01}.
Steric interactions between two squirmers and between a squirmer and a wall are taken into account by the procedure described in Appendix \ref{app:steric_interactions}.
The initial positions and orientations of the two squirmers ($i=1,2$) are
\begin{align}
  \bm{C}_{1/2}=\left(\frac{L_x}{2} \mp \frac{d_{cm}}{2},\frac{L_y}{2}, \frac{L_z}{2} \right)^T, \\
  \bm{e}_{1/2}=(\pm \cos(\alpha_0),0,\sin(\alpha_0))^T.
\end{align}
Here, $d_{cm}$ is the initial center-of-mass distance and $\alpha_0=(\pi-\theta_0)/2$, where $\theta_0$ is the inital angle between $\bm{e}_1$ and $\bm{e}_2$.
The swimming mode is chosen as $B_1=0.05 \sqrt{k_B T/m}$ and the force dipole mode $\beta \in \{-4,0,4\}$.
We choose $d_{cm}$ such that the squirmers are well separated and vary $\theta_0$. The squirmers major and minor axes are $b_x=3a$ and $b_z=6a$, respectively,  and the simulation box size is $L_x=L_z=15b_z$, and $L_y=7a$. Note that $L_y \gtrsim b_x$ which keeps the swimming orientation essentially in the $x$-$z$ plane.

Results for the mean surface-to-surface distance between squirmers $\lla d_s \rra$ and the mean alignment $\lla {\bm e}_1 \cdot {\bm e}_2 \rra = \lla \cos \theta \rra $ are shown in Fig.~\ref{fig:d_cos} for pushers, pullers, and neutral swimmers with an inital angle $\theta_0= 3 \pi/8$.
Due to the setup, the squirmers initially approach each other and collide at $tU_0/\sigma \approx 0.5$.
The (persistence) P\'eclet number $Pe=v_0/(2D_R^\perp \sigma)\approx 60$ is sufficiently high, such that the squirmer orientation has hardly changed before collision.
When the neutral swimmers collide, they initially align parallel ($\cos \theta \approx 1$ at $tU_0/\sigma \approx 1$ in Fig.~\ref{fig:d_cos}), but their trajectories start to diverge immediately thereafter. Pushers remain parallel for an extended time window, which is expected as pushers are known to attract each other \cite{goet:10a}, but at $tU_0/\sigma \approx 3$ (cf. Fig.~\ref{fig:d_cos}) their trajectories diverge as well. This is probably due to noise, since we observe several realizations where pushers remain parallel. Interestingly, pullers, which are known to repel each other when swimming in parallel\cite{goet:10a}, swim cooperatively and reach a stable orientation with $\lla \cos(\theta) \rra \approx 0.77$ shortly after they collided (at $tU_0/\sigma \approx 1$). Thereby, their cooperative swimming velocity is about $0.8 U_0$. The flow field of this stable state, determined by MPC simulations, is shown in Fig. \ref{fig:coop_flow_field}. Note that the velocity field in the swimming plane is left-right symmetric, and that there is a stagnation point in the center behind the swimmers. Figure \ref{fig:coop_flow_field} reveals that this point actually corresponds to a line normal to the walls.
\begin{figure}
\includegraphics*[width=\columnwidth]{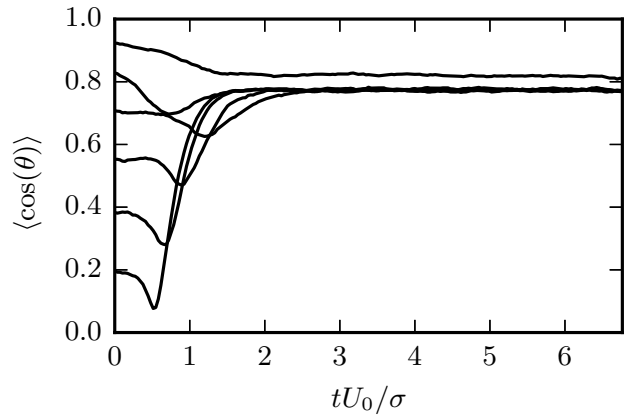}
\caption{\label{fig:initial_conditions_to_fixed_point}
  Time dependence of the average alignment $\lla {\bm e_1}\cdot {\bm e}_2 \rra = \langle\cos \theta \rangle$ of two pullers with $\beta=4$, $b_z/b_x=2$, and various initial angles $\theta_0 \in (0, \pi/2)$.
}
\end{figure}
%
Figure~\ref{fig:initial_conditions_to_fixed_point} shows that the fixed point of cooperatively swimming  pullers is reached for nearly all simulated initial conditions $\theta_0 \in (0,\pi/2)$.
Only pullers that are nearly parallel initially ($\theta_0=\pi/8$, $\cos \theta_0 \approx 0.92$ in Fig. ~\ref{fig:initial_conditions_to_fixed_point}), repel each other such that they will not reach the fixed point.
For  P\'eclet numbers $Pe < 60$,  the fixed point remains at $\lla \cos(\theta) \rra \approx 0.77$. However, it becomes more likely for the swimmers to escape (or never reach) the fixed point.

A detailed study reveals that the fixed point vanishes, when the walls are replaced by periodic boundary conditions.
This is even true when we apply three-dimensional periodic boundaries, but keep the wall potential implemented, i.e., the squirmers are still confined in a narrow slit. In addition, we studied the swim behavior of spherical squirmers. Here,  we observe diverging  trajectories for all squirmer types, i.e., pushers, neutral squirmers, and pullers. Such diverging trajectories have already been reported in Ref.~\onlinecite{goet:10} for spherical squirmers in bulk.
Hence, the stable close-by cooperative swimming of pullers is governed by the squirmer anisotropy, by the hydrodynamic interactions between them and, importantly, between pullers and confining surfaces.

This conclusion is in contrast to results presented in Ref.~\onlinecite{kyoy:15}, where a monolayer of spheroidal squirmers is considered,
with their centers and orientation vectors fixed in the same plane, however, without confining walls.
The study reports a stable cooperative motion for pullers with angles $\theta \in (0,\pi/2)$ by nearest-neighbor two-body interactions, where all angles between $0$ and $\pi/2$ are stable.
The difference to our study is that in Ref.~\onlinecite{kyoy:15} cooperative features were extracted from a simulation of many swimmers, whereas we explicitly studied two swimmers.
Furthermore our study explicitly models no-slip walls and includes thermal noise.

To shed light on the stability of the cooperative puller motion, we varied the puller strength $\beta$, the aspect ratio $b_z/b_x$, and the width of the slit $L_y$. Thereby, we started from our basic  parameter set $b_x=3a, ~b_z=6a,~ L_y=7a, $ and $\beta=4$.

With decreasing $\beta$, the stable alignment disappears, i.e., the pullers' distance increases after collision. For increasing $\beta$ the fixed point remains, but the value of $\cos \theta$ decreases, i.e., the squirmers form a larger angle.


With increasing wall separation, the fixed-point value of $\cos \theta$  decreases, i.e, the angle between the swimmers increases.
For $L_y/b_z=2$ and higher, the fixed point disappears.

An increase of the aspect ratio $b_z/b_x$ from $2$ to $3$ and $4$ increases the fixed-point value of $ \lla \cos \theta \rra$ from $0.77$ to $0.84$ and $0.88$. The more elongated shape leads to a more parallel alignment of the squirmers. The minimal value of $\beta$ required to achieve cooperative motion depends weakly on the aspect ratio. For $b_z/b_x=2, \ 3,$ and $4$ the critical values for $\beta$ are $\approx 3.7, \ 3.6,$ and $3.4$. Hence, a large aspect ratio is beneficial for cooperative swimming.


\section{Summary and conclusions}

We have introduced a spheroidal squirmer model, which comprises the swimming and force-dipole modes. It is a variation of previously proposed squirmer models. On the one hand, it includes the force-dipole mode as an extension to the model of Ref.~\onlinecite{kell:77}. On the other hand, it is an alternative approach compared to Refs.~\onlinecite{ishi:13} and \onlinecite{kyoy:15}, with the major advantage that our model allows for the analytical calculation of the flow field. In the present calculations we employed the Stokes stream function equation. Very recently a full set of solutions to Stokes' equations in spheroidal coordinates were given in Ref.~\onlinecite{feld:16}, which opens an alternative approach to derive the flow field for our choice of  boundary conditions.

Furthermore, we have presented an implementation of our spheroidal squirmer in a MPC fluid.  In contrast to other frequently employed simulation approaches, MPC includes thermal fluctuations. The comparison between the fluid flow profile of a squirmer extracted from the simulation data with the theoretical prediction yields very good agreement. As a consequence of the MPC approach with its discrete collision cells, the minor axis of the spheroid has to be larger than a few collision cells to avoid discretization effects. The analysis of the squirmer orientation correlation function shows that very good agreement between theory and simulations is already obtained for $bz=9a$ (major axis) and $b_x=3a$ (minor axis).

To shed light on the cooperative swimming motion and on near-field hydrodynamic interactions,  we investigated the collision of two spheroidal squirmers in a slit geometry.
We found stable stationary states of close-by swimming for spheroidal pullers, which is determined by  hydrodynamic interactions between the anisotropic squirmers, and, even more important, by  squirmers and surfaces.
This stationary state disappears for low puller strengths and low eccentricities.
We expect the stable close-by swimming of pullers to strongly enhance clustering in puller suspensions in thin films.

Our studies confirm that spheroidal squirmers can accurately be simulated by the MPC method. The proposed implementation opens an avenue to study collective and non-equilibrium effects in systems of  anisotropic microswimmers. Even large-scale systems can be addressed by the implementation of MPC and the squirmer dynamics on GPUs.

\section*{Acknowledgments}
We thank A. Wysocki, A. Varghese, and  B.U. Felderhof for helpful discussions. Support of this work by the DFG priority program SPP 1726 on ``Microswimmers -- from Single Particle Motion to Collective Behaviour'' is gratefully acknowledged.

\appendix

\section{Quaternion matrices}
The rotation matrix $\bm{D}$ introduced in Eq. (\ref{Eq:vb_equals_Dvs}) is given in terms of the rotation quaternion $\bm{q}$ as
\begin{align} \label{Eq:Rotation_matrix_for_RBD}
  \begin{pmatrix}
          q_0^2+q_1^2-q_2^2-q_3^2 & 2(q_1 q_2+q_0 q_3) & 2(q_1 q_3-q_0 q_2) \\
          2(q_2q_1-q_0q_3) & q_0^2-q_1^2+q_2^2-q_3^2 & 2(q_2q_3+q_0q_1)\\
          2(q_2q_1+q_0q_2) & 2(q_3q_2-q_0q_1) & q_0^2-q_1^2-q_2^2+q_3^2
          \end{pmatrix}
\end{align}
The matrix $\bm{Q}(\bm{q})$ in Eq. (\ref{Eq:qdot_equals}) is given by
\begin{align} \label{Eq:def_Q_of_q}
  \bm{Q}(\bm{q})&=\begin{pmatrix*}[r]
		      q_0 & -q_1 & -q_2 & -q_3 \\
		      q_1 & q_0 & -q_3 & q_2 \\
		      q_2 & q_3 & q_0 & -q_1 \\
		      q_3 & -q_2 & q_1 & q_0
		\end{pmatrix*}.
\end{align}

\section{Gegenbauer functions} \label{app:Gegenbauer}
For $n \geq 2$ and $x \in \mathbb{R}$ the Gegenbauer functions of the first and second kind $G_n$ and $H_n$,
are defined in terms of the Legendre functions of the first and second kind $P_n$ and $Q_n$ as \cite{wang:89,dass:08}
\begin{align}
  G_n(x)=\frac{P_{n-2}(x)-P_n(x)}{2n-1}, ~ H_n(x)=\frac{Q_{n-2}(x)-Q_n(x)}{2n-1}.
\end{align}
For $n=0,1$, they are defined as
\begin{align}
  G_0(x)=-H_1(x)=1, ~ G_1(x)=H_0(x)=-x.
\end{align}
For the reader's convenience, we give the formula for the Gegenbauer functions of the first kind for $n=2,3$, and $x \in \mathbb{R}$
\begin{align}
  G_2(x)&=\frac{1}{2}(1-x^2),\\
  G_3(x)&=\frac{1}{2}(1-x^2)x.
\end{align}
Furthermore, the Gegenbauer functions of the second kind for $n=2,3$, and $x>1$ are given by
\begin{align}
  H_2(x)&=\frac{1}{2}(1-x^2) \coth ^{-1}(x)+\frac{x}{2},\\
  H_3(x)&=\frac{1}{2}(1-x^2)x \coth ^{-1}(x)+\frac{1}{6}(3x^2-2).
\end{align}
Here, we used $\coth ^{-1}(x)= \ln ([x+1]/[x-1])/2 $.

\section{Steric interactions} \label{app:steric_interactions}

Here, we illustrate our implementation of the excluded-volume interactions between spheroids and walls following the approach provided in Ref.~\onlinecite{para:05}.

The spheroid's surface  in the laboratory frame is given by the quadratic form
\begin{align}
  1=\mathcal{A}(\bm{x}) \equiv (\bm{x}-\bm{C})^T\bm{A}(\bm{x}-\bm{C}),
\end{align}
where the orientation matrix $\bm{A}$ can be expressed as
\begin{align}
  \bm{A}=(1-\bm{e}\bm{e}^T)/b_x^2+ \bm{e}\bm{e}^T /b_z^{2}.
\end{align}
For the steric interactions, we introduce a virtual safety distance $d_v$, which is small compared to $b_x$ and $b_z$.
When computing steric interactions, we replace $b_x$ and $b_z$ by $b_x+d_v$ and $b_z+d_v$, respectively.
In this paper we used $d_v=0.05a$ for all simulations.

\subsection{Interaction between spheroids}

We introduce a repulsive interaction potential between spheroids to prevent their overlap.
The potential is given by
\begin{align} \label{Eq:potential_of_dR}
  U=4\epsilon_0 \left[ \left(\frac{\sigma_0}{d_R+\sigma_0} \right)^{12} - \left(\frac{\sigma_0}{d_R+\sigma_0} \right)^{6}\right].
\end{align}
Here, $\sigma_0$ and $\epsilon_0$ correspond to a length and energy scale, respectively. We choose $\epsilon_0=k_BT$ and $\sigma_0=2 d_v$.
The directional contact distance $d_R$ between two spheroids, with orientation matrices $\bm{A}_1$, $\bm{A}_2$ and center positions
$\bm{C}_1$, $\bm{C}_2$, is an approximation to their true distance of closest approach and is defined by
\begin{align} \label{Eq:Def_dR_of_FA1A2}
  d_R=R(1-F(\bm{A}_1,\bm{A}_2)^{-1/2})
\end{align}
Here, $\bm{R}=\bm{C}_2-\bm{C}_1$, $R=|\bm{R}|$, and $F(\bm{A}_1,\bm{A}_2)$ is the elliptic contact function, defined as \cite{para:05}
\begin{align}
  F(\bm{A}_1,\bm{A}_2)&=\max_\lambda \min_{\bm{x}} \mathcal{S}(\bm{x},\lambda)\\
  &=\max_\lambda \min_{\bm{x}} \left( \lambda \mathcal{A}_1(\bm{x})+(1-\lambda) \mathcal{A}_2(\bm{x}) \right).
\end{align}
Minimization with respect to $\bm{x}$ demands $\bm{\nabla} \mathcal{S}(\bm{x},\lambda)=0$, and hence,
\begin{align}
  \bm{x}(\lambda)=\left\{\lambda \bm{A}_1+(1-\lambda) \bm{A}_2 \right\}^{-1}  \left\{ \lambda \bm{A}_1 \bm{C}_1 +(1-\lambda) \bm{A}_2 \bm{C}_2 \right\}.
\end{align}
The critical value $\lambda=\lambda_c$ that maximizes $\mathcal{S}(\bm{x}(\lambda),\lambda)$ can be found by the root finding problem
\begin{align} \label{Eq:A1_minus_A2_equals_zero}
  \mathcal{A}_1(\bm{x}(\lambda))-\mathcal{A}_2(\bm{x}(\lambda))=0.
\end{align}
We implement Brent's  root finding approach \cite{pres:07}.
The forces and torques arising from the potential (\ref{Eq:potential_of_dR}) can be calculated analytically and are given by \cite{para:05}
\footnote{Note that Eq.~(54) of Ref.~\onlinecite{para:05} contains a typographical error. The factor 24 needs to be replaced by 12.}
\begin{align} \nonumber
  \bm{F}_1 = & \frac{24 \epsilon_0}{\sigma_0} \left[ 2 \left( \frac{\sigma_0}{d_R+\sigma_0}\right)^{13} - \left( \frac{\sigma_0}{d_R+\sigma_0}\right)^{7}\right] \\ & \times \left( \frac{\bm{R}}{R}(F^{-1/2}-1) -\frac{R}{2}F^{-3/2} \bm{X}_c\right),
\end{align}
and
\begin{align}
  \bm{T}_1 = & -\frac{12 R \epsilon_0}{\sigma_0} \left[ 2 \left( \frac{\sigma_0}{d_R+\sigma_0}\right)^{13} - \left( \frac{\sigma_0}{d_R+\sigma_0}\right)^{7}\right] \\ & \times F^{-3/2} (\bm{x}_c-\bm{C})\times \bm{X}_c
\end{align}
for the first spheroid, where $ \bm{X}_c=2\lambda_c \bm{A}_1 (\bm{x}_c-\bm{C}_1)$.
The force and torque on the second spheroid follow by Newton's action-reaction law, namely
\begin{align}
  \bm{F}_2&=-\bm{F}_1, \\
  \bm{T}_2&=-\bm{T}_1+\bm{R}\times \bm{F}_1.
\end{align}
We restrict ourselves to short-rang repulsive interactions by setting the potential $U$ to a constant value for $d_R>(\sqrt[6]{2}-1)\sigma_0$, which implies that $\bm{F}_1$ and $ \bm{T}_1$ are  zero for this range of $d_R$ values.
Note that an upper bound to $d_R$ is $R-2b_z$, which means that two spheroids will not interact if $R>2b_z+(\sqrt[6]{2}-1)\sigma_0$.
This inequality is checked before a numerical calculation of $d_R$ is employed.

\subsection{Interaction between a spheroid and a wall} \label{sec:wall_repulsion}

We assume that two parallel walls are positioned at $y=0,L_y$, which---taking into account the safety distance $d_v$---results in the effective wall positions $y=d_v$ and $L_y-d_v$.
We propose an interaction between a spheroid and a wall in the style of the spheroid-spheroid interaction presented in Ref.~ \onlinecite{para:05}.
First, we find the point $\bm{x}$ on the spheroid's surface that is closest to a wall.
For the wall at $y=d_v$, this is achieved by
minimizing the height $h(\bm{x})=\bm{e}_y \cdot \bm{x}-d_v$ under the constraint $\mathcal{A}(\bm{x})=1$. Using the method of Lagrange multipliers, we have
to minimize $\Lambda(\bm{x},\lambda)=h(\bm{x})+\lambda(\mathcal{A}(\bm{x})-1)$. The necessary condition for a minimum $\partial \Lambda/ \partial \bm{x} = 0$ yields
\begin{align}
  \bm{e}_y+ \lambda \bm{\nabla} \mathcal{A}(\bm{x})=\bm{e}_y+ 2 \lambda \bm{A}(\bm{x}-\bm{C}) = 0,
\end{align}
and hence,
\begin{align} \label{Eq:x_equals_C-stuff_over_2_lambda}
  \bm{x}=\bm{C} - \bm{A}^{-1} \bm{e}_y /(2 \lambda).
\end{align}
Substitution of Eq. (\ref{Eq:x_equals_C-stuff_over_2_lambda}) into $\mathcal{A}(\bm{x})=1$ yields
\begin{align}
  \lambda=\pm \sqrt{(\bm{A}^{-1})_{yy}}/2
\end{align}
Finally, we obtain the point closest to the wall as
\begin{align}
  \bm{x}=\bm{C} \mp (\bm{A}^{-1} \bm{e}_y)/\sqrt{(\bm{A}^{-1})_{yy}}.
\end{align}
Here, the minus sign has to be chosen, which can be visualized by the example of a sphere of radius $R$, for which $\bm{A}=R^{-2} \bm{1}$.
This finally yields the height
\begin{align}
  h=C_y-d_v-\sqrt{(\bm{A}^{-1})_{yy}} .
\end{align}
We employ the Lennard-Jones potential
\begin{align}
  U_w=
  4\epsilon_0 \left[ \left(\frac{\sigma_0}{h+\sigma_0} \right)^{12} - \left(\frac{\sigma_0}{h+\sigma_0} \right)^{6}\right]
\end{align}
for a repulsive wall, and $U_w$ assumes a constant value for all $h \geq (\sqrt[6]{2}-1)\sigma_0$.
We can derive the force $F_\alpha=-\partial U_w/ \partial C_\alpha$ and torque $T_\alpha=-\partial U_w/ \partial \psi_\alpha$ acting on the spheroid analytically.
For the force, we find
\begin{align}
  \bm{F}&=-\frac{\partial U_w}{\partial h} \frac{\partial h}{\partial C_y} \bm{e}_y\\
  &=- 24 \frac{\epsilon_0}{\sigma_0} \left[ 2 \left( \frac{\sigma_0}{h+\sigma_0} \right)^{13} - \left( \frac{\sigma_0}{h+\sigma_0} \right)^7 \right]  \bm{e}_y
\end{align}
 and for the torque
\begin{align}
  T_\alpha=-\frac{\partial U_w}{\partial h} \frac{\partial h}{\partial \psi_\alpha} ,
\end{align}
with
\begin{align}
  \frac{\partial h}{\partial \psi_\alpha}=\frac{1}{\sqrt{(\bm{A}^{-1})_{yy}}} \left( \delta_{\alpha x} (\bm{A}^{-1})_{yz} - \delta_{\alpha z}  (\bm{A}^{-1})_{yx}  \right) .
\end{align}
Here, we use the relation
\begin{align}
  \frac{d}{dt} \bm{B}^{-1} = -\bm{B}^{-1} \left(\frac{d}{dt} \bm{B} \right)\bm{B}^{-1} ,
\end{align}
which holds for an invertible matrix $\bm{B} = \bm{B}(t)$ depending on a scalar parameter $t$, and Eq. (C9) from Ref.~ \onlinecite{para:05}.

For the wall at $y=L_y-d_v$, we have to minimize $h(\bm{x})=L_y-d_v-\bm{e}_y\cdot \bm{x}$, with $\bm{x}$ on the spheroid's surface. This yields
\begin{align}
  \bm{x}=\bm{C} + \bm{A}^{-1} \bm{e_y} \left[ \left(\bm{A}^{-1} \right)_{yy}\right]^{-1/2} .
\end{align}
The formulas for torque and force do not change, except that we have to insert $h=L_y-d_v-C_y-\sqrt{(\bm{A}^{-1})_{yy}}$ and need to change the sign of the force.


\footnotesize{
\bibliographystyle{rsc} 

\begin{mcitethebibliography}{89}
\providecommand*{\natexlab}[1]{#1}
\providecommand*{\mciteSetBstSublistMode}[1]{}
\providecommand*{\mciteSetBstMaxWidthForm}[2]{}
\providecommand*{\mciteBstWouldAddEndPuncttrue}
  {\def\EndOfBibitem{\unskip.}}
\providecommand*{\mciteBstWouldAddEndPunctfalse}
  {\let\EndOfBibitem\relax}
\providecommand*{\mciteSetBstMidEndSepPunct}[3]{}
\providecommand*{\mciteSetBstSublistLabelBeginEnd}[3]{}
\providecommand*{\EndOfBibitem}{}
\mciteSetBstSublistMode{f}
\mciteSetBstMaxWidthForm{subitem}
{(\emph{\alph{mcitesubitemcount}})}
\mciteSetBstSublistLabelBeginEnd{\mcitemaxwidthsubitemform\space}
{\relax}{\relax}

\bibitem[Elgeti \emph{et~al.}(2015)Elgeti, Winkler, and Gompper]{elge:15}
J.~Elgeti, R.~G. Winkler and G.~Gompper, \emph{Reports on Progress in Physics},
  2015, \textbf{78}, 056601\relax
\mciteBstWouldAddEndPuncttrue
\mciteSetBstMidEndSepPunct{\mcitedefaultmidpunct}
{\mcitedefaultendpunct}{\mcitedefaultseppunct}\relax
\EndOfBibitem
\bibitem[Vicsek and Zafeiris(2012)]{vics:12}
T.~Vicsek and A.~Zafeiris, \emph{Phys. Rep.}, 2012, \textbf{517}, 71\relax
\mciteBstWouldAddEndPuncttrue
\mciteSetBstMidEndSepPunct{\mcitedefaultmidpunct}
{\mcitedefaultendpunct}{\mcitedefaultseppunct}\relax
\EndOfBibitem
\bibitem[Copeland and Weibel(2009)]{cope:09}
M.~F. Copeland and D.~B. Weibel, \emph{Soft Matter}, 2009, \textbf{5},
  1174\relax
\mciteBstWouldAddEndPuncttrue
\mciteSetBstMidEndSepPunct{\mcitedefaultmidpunct}
{\mcitedefaultendpunct}{\mcitedefaultseppunct}\relax
\EndOfBibitem
\bibitem[Darnton \emph{et~al.}(2010)Darnton, Turner, Rojevsky, and
  Berg]{darn:10}
N.~C. Darnton, L.~Turner, S.~Rojevsky and H.~C. Berg, \emph{Biophys. J.}, 2010,
  \textbf{98}, 2082\relax
\mciteBstWouldAddEndPuncttrue
\mciteSetBstMidEndSepPunct{\mcitedefaultmidpunct}
{\mcitedefaultendpunct}{\mcitedefaultseppunct}\relax
\EndOfBibitem
\bibitem[Kearns(2010)]{kear:10}
D.~B. Kearns, \emph{Nat. Rev. Microbiol.}, 2010, \textbf{8}, 634--644\relax
\mciteBstWouldAddEndPuncttrue
\mciteSetBstMidEndSepPunct{\mcitedefaultmidpunct}
{\mcitedefaultendpunct}{\mcitedefaultseppunct}\relax
\EndOfBibitem
\bibitem[Drescher \emph{et~al.}(2011)Drescher, Dunkel, Cisneros, Ganguly, and
  Goldstein]{dres:11}
K.~Drescher, J.~Dunkel, L.~H. Cisneros, S.~Ganguly and R.~E. Goldstein,
  \emph{Proc. Natl. Acad. Sci. USA}, 2011, \textbf{10940}, 108\relax
\mciteBstWouldAddEndPuncttrue
\mciteSetBstMidEndSepPunct{\mcitedefaultmidpunct}
{\mcitedefaultendpunct}{\mcitedefaultseppunct}\relax
\EndOfBibitem
\bibitem[Partridge and Harshey(2013)]{part:13}
J.~D. Partridge and R.~M. Harshey, \emph{J. Bacteriol.}, 2013, \textbf{195},
  909\relax
\mciteBstWouldAddEndPuncttrue
\mciteSetBstMidEndSepPunct{\mcitedefaultmidpunct}
{\mcitedefaultendpunct}{\mcitedefaultseppunct}\relax
\EndOfBibitem
\bibitem[Bialk{\'e} \emph{et~al.}(2012)Bialk{\'e}, Speck, and
  L{\"o}wen]{bial:12}
J.~Bialk{\'e}, T.~Speck and H.~L{\"o}wen, \emph{Phys. Rev. Lett.}, 2012,
  \textbf{108}, 168301\relax
\mciteBstWouldAddEndPuncttrue
\mciteSetBstMidEndSepPunct{\mcitedefaultmidpunct}
{\mcitedefaultendpunct}{\mcitedefaultseppunct}\relax
\EndOfBibitem
\bibitem[Buttinoni \emph{et~al.}(2013)Buttinoni, Bialk{\'e}, K{\"u}mmel,
  L{\"o}wen, Bechinger, and Speck]{butt:13}
I.~Buttinoni, J.~Bialk{\'e}, F.~K{\"u}mmel, H.~L{\"o}wen, C.~Bechinger and
  T.~Speck, \emph{Phys. Rev. Lett.}, 2013, \textbf{110}, 238301\relax
\mciteBstWouldAddEndPuncttrue
\mciteSetBstMidEndSepPunct{\mcitedefaultmidpunct}
{\mcitedefaultendpunct}{\mcitedefaultseppunct}\relax
\EndOfBibitem
\bibitem[Mognetti \emph{et~al.}(2013)Mognetti, {\v S}ari{\'c},
  Angioletti-Uberti, Cacciuto, Valeriani, and Frenkel]{mogn:13}
B.~M. Mognetti, A.~{\v S}ari{\'c}, S.~Angioletti-Uberti, A.~Cacciuto,
  C.~Valeriani and D.~Frenkel, \emph{Phys. Rev. Lett.}, 2013, \textbf{111},
  245702\relax
\mciteBstWouldAddEndPuncttrue
\mciteSetBstMidEndSepPunct{\mcitedefaultmidpunct}
{\mcitedefaultendpunct}{\mcitedefaultseppunct}\relax
\EndOfBibitem
\bibitem[Theurkauff \emph{et~al.}(2012)Theurkauff, Cottin-Bizonne, Palacci,
  Ybert, and Bocquet]{theu:12}
I.~Theurkauff, C.~Cottin-Bizonne, J.~Palacci, C.~Ybert and L.~Bocquet,
  \emph{Phys. Rev. Lett.}, 2012, \textbf{108}, 268303\relax
\mciteBstWouldAddEndPuncttrue
\mciteSetBstMidEndSepPunct{\mcitedefaultmidpunct}
{\mcitedefaultendpunct}{\mcitedefaultseppunct}\relax
\EndOfBibitem
\bibitem[Fily \emph{et~al.}(2014)Fily, Henkes, and Marchetti]{fily:14.1}
Y.~Fily, S.~Henkes and M.~C. Marchetti, \emph{Soft Matter}, 2014, \textbf{10},
  2132\relax
\mciteBstWouldAddEndPuncttrue
\mciteSetBstMidEndSepPunct{\mcitedefaultmidpunct}
{\mcitedefaultendpunct}{\mcitedefaultseppunct}\relax
\EndOfBibitem
\bibitem[Yang \emph{et~al.}(2014)Yang, Manning, and Marchetti]{yang:14.2}
X.~Yang, M.~L. Manning and M.~C. Marchetti, \emph{Soft Matter}, 2014,
  \textbf{10}, 6477\relax
\mciteBstWouldAddEndPuncttrue
\mciteSetBstMidEndSepPunct{\mcitedefaultmidpunct}
{\mcitedefaultendpunct}{\mcitedefaultseppunct}\relax
\EndOfBibitem
\bibitem[Stenhammar \emph{et~al.}(2014)Stenhammar, Marenduzzo, Allen, and
  Cates]{sten:14}
J.~Stenhammar, D.~Marenduzzo, R.~J. Allen and M.~E. Cates, \emph{Soft Matter},
  2014, \textbf{10}, 1489\relax
\mciteBstWouldAddEndPuncttrue
\mciteSetBstMidEndSepPunct{\mcitedefaultmidpunct}
{\mcitedefaultendpunct}{\mcitedefaultseppunct}\relax
\EndOfBibitem
\bibitem[Fily \emph{et~al.}(2014)Fily, Baskaran, and Hagan]{fily:14}
Y.~Fily, A.~Baskaran and M.~F. Hagan, \emph{Soft Matter}, 2014, \textbf{10},
  5609\relax
\mciteBstWouldAddEndPuncttrue
\mciteSetBstMidEndSepPunct{\mcitedefaultmidpunct}
{\mcitedefaultendpunct}{\mcitedefaultseppunct}\relax
\EndOfBibitem
\bibitem[Redner \emph{et~al.}(2013)Redner, Hagan, and Baskaran]{redn:13}
G.~S. Redner, M.~F. Hagan and A.~Baskaran, \emph{Phys. Rev. Lett.}, 2013,
  \textbf{110}, 055701\relax
\mciteBstWouldAddEndPuncttrue
\mciteSetBstMidEndSepPunct{\mcitedefaultmidpunct}
{\mcitedefaultendpunct}{\mcitedefaultseppunct}\relax
\EndOfBibitem
\bibitem[Fily and Marchetti(2012)]{fily:12}
Y.~Fily and M.~C. Marchetti, \emph{Phys. Rev. Lett.}, 2012, \textbf{108},
  235702\relax
\mciteBstWouldAddEndPuncttrue
\mciteSetBstMidEndSepPunct{\mcitedefaultmidpunct}
{\mcitedefaultendpunct}{\mcitedefaultseppunct}\relax
\EndOfBibitem
\bibitem[Gro{\ss}mann \emph{et~al.}(2012)Gro{\ss}mann, Schimansky-Geier, and
  Romanczuk]{gros:12}
R.~Gro{\ss}mann, L.~Schimansky-Geier and P.~Romanczuk, \emph{New J. Phys.},
  2012, \textbf{14}, 073033\relax
\mciteBstWouldAddEndPuncttrue
\mciteSetBstMidEndSepPunct{\mcitedefaultmidpunct}
{\mcitedefaultendpunct}{\mcitedefaultseppunct}\relax
\EndOfBibitem
\bibitem[Lobaskin and Romenskyy(2013)]{loba:13}
V.~Lobaskin and M.~Romenskyy, \emph{Phys. Rev. E}, 2013, \textbf{87},
  052135\relax
\mciteBstWouldAddEndPuncttrue
\mciteSetBstMidEndSepPunct{\mcitedefaultmidpunct}
{\mcitedefaultendpunct}{\mcitedefaultseppunct}\relax
\EndOfBibitem
\bibitem[Z{\"o}ttl and Stark(2014)]{zoet:14}
A.~Z{\"o}ttl and H.~Stark, \emph{Phys. Rev. Lett.}, 2014, \textbf{112},
  118101\relax
\mciteBstWouldAddEndPuncttrue
\mciteSetBstMidEndSepPunct{\mcitedefaultmidpunct}
{\mcitedefaultendpunct}{\mcitedefaultseppunct}\relax
\EndOfBibitem
\bibitem[Wysocki \emph{et~al.}(2014)Wysocki, Winkler, and Gompper]{wyso:14}
A.~Wysocki, R.~G. Winkler and G.~Gompper, \emph{EPL (Europhysics Letters)},
  2014, \textbf{105}, 48004\relax
\mciteBstWouldAddEndPuncttrue
\mciteSetBstMidEndSepPunct{\mcitedefaultmidpunct}
{\mcitedefaultendpunct}{\mcitedefaultseppunct}\relax
\EndOfBibitem
\bibitem[Peruani \emph{et~al.}(2006)Peruani, Deutsch, and B{\"a}r]{peru:06}
F.~Peruani, A.~Deutsch and M.~B{\"a}r, \emph{Phys. Rev. E}, 2006, \textbf{74},
  030904\relax
\mciteBstWouldAddEndPuncttrue
\mciteSetBstMidEndSepPunct{\mcitedefaultmidpunct}
{\mcitedefaultendpunct}{\mcitedefaultseppunct}\relax
\EndOfBibitem
\bibitem[Yang \emph{et~al.}(2010)Yang, Marceau, and Gompper]{yang:10}
Y.~Yang, V.~Marceau and G.~Gompper, \emph{Phys. Rev. E}, 2010, \textbf{82},
  031904\relax
\mciteBstWouldAddEndPuncttrue
\mciteSetBstMidEndSepPunct{\mcitedefaultmidpunct}
{\mcitedefaultendpunct}{\mcitedefaultseppunct}\relax
\EndOfBibitem
\bibitem[Elgeti and Gompper(2009)]{elge:09}
J.~Elgeti and G.~Gompper, \emph{EPL (Europhysics Letters)}, 2009, \textbf{85},
  38002\relax
\mciteBstWouldAddEndPuncttrue
\mciteSetBstMidEndSepPunct{\mcitedefaultmidpunct}
{\mcitedefaultendpunct}{\mcitedefaultseppunct}\relax
\EndOfBibitem
\bibitem[Berke \emph{et~al.}(2008)Berke, Turner, Berg, and Lauga]{berk:08}
A.~P. Berke, L.~Turner, H.~C. Berg and E.~Lauga, \emph{Phys. Rev. Lett.}, 2008,
  \textbf{101}, 038102\relax
\mciteBstWouldAddEndPuncttrue
\mciteSetBstMidEndSepPunct{\mcitedefaultmidpunct}
{\mcitedefaultendpunct}{\mcitedefaultseppunct}\relax
\EndOfBibitem
\bibitem[Lauga and Powers(2009)]{laug:09}
E.~Lauga and T.~R. Powers, \emph{Reports on Progress in Physics}, 2009,
  \textbf{72}, 096601\relax
\mciteBstWouldAddEndPuncttrue
\mciteSetBstMidEndSepPunct{\mcitedefaultmidpunct}
{\mcitedefaultendpunct}{\mcitedefaultseppunct}\relax
\EndOfBibitem
\bibitem[Lauga \emph{et~al.}(2006)Lauga, DiLuzio, Whitesides, and
  Stone]{laug:06}
E.~Lauga, W.~R. DiLuzio, G.~M. Whitesides and H.~A. Stone, \emph{Biophys. J.},
  2006, \textbf{90}, 400\relax
\mciteBstWouldAddEndPuncttrue
\mciteSetBstMidEndSepPunct{\mcitedefaultmidpunct}
{\mcitedefaultendpunct}{\mcitedefaultseppunct}\relax
\EndOfBibitem
\bibitem[Hu \emph{et~al.}(2015)Hu, Wysocki, Winkler, and Gompper]{hu:15}
J.~Hu, A.~Wysocki, R.~G. Winkler and G.~Gompper, \emph{Scientific Reports},
  2015, \textbf{5}, 9586\relax
\mciteBstWouldAddEndPuncttrue
\mciteSetBstMidEndSepPunct{\mcitedefaultmidpunct}
{\mcitedefaultendpunct}{\mcitedefaultseppunct}\relax
\EndOfBibitem
\bibitem[Di~Leonardo \emph{et~al.}(2011)Di~Leonardo, Dell'Arciprete, Angelani,
  and Iebba]{dile:11}
R.~Di~Leonardo, D.~Dell'Arciprete, L.~Angelani and V.~Iebba, \emph{Phys. Rev.
  Lett.}, 2011, \textbf{106}, 038101\relax
\mciteBstWouldAddEndPuncttrue
\mciteSetBstMidEndSepPunct{\mcitedefaultmidpunct}
{\mcitedefaultendpunct}{\mcitedefaultseppunct}\relax
\EndOfBibitem
\bibitem[Lemelle \emph{et~al.}(2013)Lemelle, Palierne, Chatre, Vaillant, and
  Place]{leme:13}
L.~Lemelle, J.-F. Palierne, E.~Chatre, C.~Vaillant and C.~Place, \emph{Soft
  Matter}, 2013, \textbf{9}, 9759\relax
\mciteBstWouldAddEndPuncttrue
\mciteSetBstMidEndSepPunct{\mcitedefaultmidpunct}
{\mcitedefaultendpunct}{\mcitedefaultseppunct}\relax
\EndOfBibitem
\bibitem[Lighthill(1952)]{ligh:52}
M.~J. Lighthill, \emph{Commun. Pure Appl. Math.}, 1952, \textbf{5},
  109--118\relax
\mciteBstWouldAddEndPuncttrue
\mciteSetBstMidEndSepPunct{\mcitedefaultmidpunct}
{\mcitedefaultendpunct}{\mcitedefaultseppunct}\relax
\EndOfBibitem
\bibitem[Blake(1971)]{blak:71}
J.~R. Blake, \emph{J. Fluid Mech.}, 1971, \textbf{46}, 199--208\relax
\mciteBstWouldAddEndPuncttrue
\mciteSetBstMidEndSepPunct{\mcitedefaultmidpunct}
{\mcitedefaultendpunct}{\mcitedefaultseppunct}\relax
\EndOfBibitem
\bibitem[Howse \emph{et~al.}(2007)Howse, Jones, Ryan, Gough, Vafabakhsh, and
  Golestanian]{hows:07}
J.~R. Howse, R.~A.~L. Jones, A.~J. Ryan, T.~Gough, R.~Vafabakhsh and
  R.~Golestanian, \emph{Phys. Rev. Lett.}, 2007, \textbf{99}, 048102\relax
\mciteBstWouldAddEndPuncttrue
\mciteSetBstMidEndSepPunct{\mcitedefaultmidpunct}
{\mcitedefaultendpunct}{\mcitedefaultseppunct}\relax
\EndOfBibitem
\bibitem[Erbe \emph{et~al.}(2008)Erbe, Zientara, Baraban, Kreidler, and
  Leiderer]{erbe:08}
A.~Erbe, M.~Zientara, L.~Baraban, C.~Kreidler and P.~Leiderer, \emph{J. Phys.:
  Condens. Matter}, 2008, \textbf{20}, 404215\relax
\mciteBstWouldAddEndPuncttrue
\mciteSetBstMidEndSepPunct{\mcitedefaultmidpunct}
{\mcitedefaultendpunct}{\mcitedefaultseppunct}\relax
\EndOfBibitem
\bibitem[Volpe \emph{et~al.}(2011)Volpe, Buttinoni, Vogt, K{\"u}mmerer, and
  Bechinger]{volp:11}
G.~Volpe, I.~Buttinoni, D.~Vogt, H.~J. K{\"u}mmerer and C.~Bechinger,
  \emph{Soft Matter}, 2011, \textbf{7}, 8810\relax
\mciteBstWouldAddEndPuncttrue
\mciteSetBstMidEndSepPunct{\mcitedefaultmidpunct}
{\mcitedefaultendpunct}{\mcitedefaultseppunct}\relax
\EndOfBibitem
\bibitem[Llopis and Pagonabarraga(2010)]{llop:10}
I.~Llopis and I.~Pagonabarraga, \emph{J. Non-Newtonian Fluid Mech.}, 2010,
  \textbf{165}, 946\relax
\mciteBstWouldAddEndPuncttrue
\mciteSetBstMidEndSepPunct{\mcitedefaultmidpunct}
{\mcitedefaultendpunct}{\mcitedefaultseppunct}\relax
\EndOfBibitem
\bibitem[G{\"o}tze and Gompper(2010)]{goet:10a}
I.~O. G{\"o}tze and G.~Gompper, \emph{EPL (Europhysics Letters)}, 2010,
  \textbf{92}, 64003\relax
\mciteBstWouldAddEndPuncttrue
\mciteSetBstMidEndSepPunct{\mcitedefaultmidpunct}
{\mcitedefaultendpunct}{\mcitedefaultseppunct}\relax
\EndOfBibitem
\bibitem[Ishikawa and Hota(2006)]{ishi:06}
T.~Ishikawa and M.~Hota, \emph{J. Exp. Biol.}, 2006, \textbf{209},
  4452--4463\relax
\mciteBstWouldAddEndPuncttrue
\mciteSetBstMidEndSepPunct{\mcitedefaultmidpunct}
{\mcitedefaultendpunct}{\mcitedefaultseppunct}\relax
\EndOfBibitem
\bibitem[Ishikawa and Pedley(2007)]{ishi:07}
T.~Ishikawa and T.~J. Pedley, \emph{J. Fluid Mech.}, 2007, \textbf{588},
  399--435\relax
\mciteBstWouldAddEndPuncttrue
\mciteSetBstMidEndSepPunct{\mcitedefaultmidpunct}
{\mcitedefaultendpunct}{\mcitedefaultseppunct}\relax
\EndOfBibitem
\bibitem[Evans \emph{et~al.}(2011)Evans, Ishikawa, Yamaguchi, and
  Lauga]{evan:11}
A.~A. Evans, T.~Ishikawa, T.~Yamaguchi and E.~Lauga, \emph{Physics of Fluids
  (1994-present)}, 2011, \textbf{23}, 111702\relax
\mciteBstWouldAddEndPuncttrue
\mciteSetBstMidEndSepPunct{\mcitedefaultmidpunct}
{\mcitedefaultendpunct}{\mcitedefaultseppunct}\relax
\EndOfBibitem
\bibitem[Alarc{\'o}n and Pagonabarraga(2013)]{alar:13}
F.~Alarc{\'o}n and I.~Pagonabarraga, \emph{J. Mol. Liq.}, 2013, \textbf{185},
  56--61\relax
\mciteBstWouldAddEndPuncttrue
\mciteSetBstMidEndSepPunct{\mcitedefaultmidpunct}
{\mcitedefaultendpunct}{\mcitedefaultseppunct}\relax
\EndOfBibitem
\bibitem[Molina \emph{et~al.}(2013)Molina, Nakayama, and Yamamoto]{moli:13}
J.~J. Molina, Y.~Nakayama and R.~Yamamoto, \emph{Soft Matter}, 2013,
  \textbf{9}, 4923\relax
\mciteBstWouldAddEndPuncttrue
\mciteSetBstMidEndSepPunct{\mcitedefaultmidpunct}
{\mcitedefaultendpunct}{\mcitedefaultseppunct}\relax
\EndOfBibitem
\bibitem[Ishikawa and Pedley(2008)]{ishi:08}
T.~Ishikawa and T.~J. Pedley, \emph{Phys. Rev. Lett.}, 2008, \textbf{100},
  088103\relax
\mciteBstWouldAddEndPuncttrue
\mciteSetBstMidEndSepPunct{\mcitedefaultmidpunct}
{\mcitedefaultendpunct}{\mcitedefaultseppunct}\relax
\EndOfBibitem
\bibitem[Ishimoto and Gaffney(2013)]{ishi:13}
K.~Ishimoto and E.~A. Gaffney, \emph{Phys. Rev. E}, 2013, \textbf{88},
  062702\relax
\mciteBstWouldAddEndPuncttrue
\mciteSetBstMidEndSepPunct{\mcitedefaultmidpunct}
{\mcitedefaultendpunct}{\mcitedefaultseppunct}\relax
\EndOfBibitem
\bibitem[Keller and Wu(1977)]{kell:77}
S.~R. Keller and T.~Y. Wu, \emph{J. Fluid Mech.}, 1977, \textbf{80},
  259--278\relax
\mciteBstWouldAddEndPuncttrue
\mciteSetBstMidEndSepPunct{\mcitedefaultmidpunct}
{\mcitedefaultendpunct}{\mcitedefaultseppunct}\relax
\EndOfBibitem
\bibitem[Li and Ardekani(2014)]{li:14}
G.-J. Li and A.~M. Ardekani, \emph{Phys. Rev. E}, 2014, \textbf{90},
  013010\relax
\mciteBstWouldAddEndPuncttrue
\mciteSetBstMidEndSepPunct{\mcitedefaultmidpunct}
{\mcitedefaultendpunct}{\mcitedefaultseppunct}\relax
\EndOfBibitem
\bibitem[Kyoya \emph{et~al.}(2015)Kyoya, Matsunaga, Imai, Omori, and
  Ishikawa]{kyoy:15}
K.~Kyoya, D.~Matsunaga, Y.~Imai, T.~Omori and T.~Ishikawa, \emph{Phys. Rev. E},
  2015, \textbf{92}, 063027\relax
\mciteBstWouldAddEndPuncttrue
\mciteSetBstMidEndSepPunct{\mcitedefaultmidpunct}
{\mcitedefaultendpunct}{\mcitedefaultseppunct}\relax
\EndOfBibitem
\bibitem[Spagnolie and Lauga(2012)]{spag:12}
S.~E. Spagnolie and E.~Lauga, \emph{J. Fluid Mech.}, 2012, \textbf{700},
  105--147\relax
\mciteBstWouldAddEndPuncttrue
\mciteSetBstMidEndSepPunct{\mcitedefaultmidpunct}
{\mcitedefaultendpunct}{\mcitedefaultseppunct}\relax
\EndOfBibitem
\bibitem[Z{\"o}ttl and Stark(2012)]{zoet:12}
A.~Z{\"o}ttl and H.~Stark, \emph{Phys. Rev. Lett.}, 2012, \textbf{108},
  218104\relax
\mciteBstWouldAddEndPuncttrue
\mciteSetBstMidEndSepPunct{\mcitedefaultmidpunct}
{\mcitedefaultendpunct}{\mcitedefaultseppunct}\relax
\EndOfBibitem
\bibitem[Pagonabarraga and Llopis(2013)]{pago:13}
I.~Pagonabarraga and I.~Llopis, \emph{Soft Matter}, 2013, \textbf{9},
  7174--7184\relax
\mciteBstWouldAddEndPuncttrue
\mciteSetBstMidEndSepPunct{\mcitedefaultmidpunct}
{\mcitedefaultendpunct}{\mcitedefaultseppunct}\relax
\EndOfBibitem
\bibitem[Delmotte \emph{et~al.}(2015)Delmotte, Keaveny, Plourabou{\'e}, and
  Climent]{delm:15}
B.~Delmotte, E.~E. Keaveny, F.~Plourabou{\'e} and E.~Climent, \emph{J. Comput.
  Phys.}, 2015, \textbf{302}, 524--547\relax
\mciteBstWouldAddEndPuncttrue
\mciteSetBstMidEndSepPunct{\mcitedefaultmidpunct}
{\mcitedefaultendpunct}{\mcitedefaultseppunct}\relax
\EndOfBibitem
\bibitem[Malevanets and Kapral(1999)]{male:99}
A.~Malevanets and R.~Kapral, \emph{J. Chem. Phys.}, 1999, \textbf{110},
  8605--8613\relax
\mciteBstWouldAddEndPuncttrue
\mciteSetBstMidEndSepPunct{\mcitedefaultmidpunct}
{\mcitedefaultendpunct}{\mcitedefaultseppunct}\relax
\EndOfBibitem
\bibitem[Kapral(2008)]{kapr:08}
R.~Kapral, \emph{Adv. Chem. Phys.}, 2008, \textbf{140}, 89--146\relax
\mciteBstWouldAddEndPuncttrue
\mciteSetBstMidEndSepPunct{\mcitedefaultmidpunct}
{\mcitedefaultendpunct}{\mcitedefaultseppunct}\relax
\EndOfBibitem
\bibitem[Gompper \emph{et~al.}(2009)Gompper, Ihle, Kroll, and Winkler]{gomp:09}
G.~Gompper, T.~Ihle, D.~M. Kroll and R.~G. Winkler, in \emph{Advanced
  {Computer} {Simulation} {Approaches} for {Soft} {Matter} {Sciences} {III}},
  ed. P.~C. Holm and P.~K. Kremer, Springer Berlin Heidelberg, 2009, pp.
  1--87\relax
\mciteBstWouldAddEndPuncttrue
\mciteSetBstMidEndSepPunct{\mcitedefaultmidpunct}
{\mcitedefaultendpunct}{\mcitedefaultseppunct}\relax
\EndOfBibitem
\bibitem[T{\"u}zel \emph{et~al.}(2006)T{\"u}zel, Ihle, and Kroll]{tuez:06}
E.~T{\"u}zel, T.~Ihle and D.~M. Kroll, \emph{Phys. Rev. E}, 2006, \textbf{74},
  056702\relax
\mciteBstWouldAddEndPuncttrue
\mciteSetBstMidEndSepPunct{\mcitedefaultmidpunct}
{\mcitedefaultendpunct}{\mcitedefaultseppunct}\relax
\EndOfBibitem
\bibitem[Huang \emph{et~al.}(2012)Huang, Gompper, and Winkler]{huan:12}
C.-C. Huang, G.~Gompper and R.~G. Winkler, \emph{Phys. Rev. E}, 2012,
  \textbf{86}, 056711\relax
\mciteBstWouldAddEndPuncttrue
\mciteSetBstMidEndSepPunct{\mcitedefaultmidpunct}
{\mcitedefaultendpunct}{\mcitedefaultseppunct}\relax
\EndOfBibitem
\bibitem[Reigh \emph{et~al.}(2012)Reigh, Winkler, and Gompper]{reig:12}
S.~Y. Reigh, R.~G. Winkler and G.~Gompper, \emph{Soft Matter}, 2012,
  \textbf{8}, 4363--4372\relax
\mciteBstWouldAddEndPuncttrue
\mciteSetBstMidEndSepPunct{\mcitedefaultmidpunct}
{\mcitedefaultendpunct}{\mcitedefaultseppunct}\relax
\EndOfBibitem
\bibitem[Reigh \emph{et~al.}(2013)Reigh, Winkler, and Gompper]{reig:13}
S.~Y. Reigh, R.~G. Winkler and G.~Gompper, \emph{{PLoS} {ONE}}, 2013,
  \textbf{8}, e70868\relax
\mciteBstWouldAddEndPuncttrue
\mciteSetBstMidEndSepPunct{\mcitedefaultmidpunct}
{\mcitedefaultendpunct}{\mcitedefaultseppunct}\relax
\EndOfBibitem
\bibitem[R{\"u}ckner and Kapral(2007)]{ruec:07}
G.~R{\"u}ckner and R.~Kapral, \emph{Phys. Rev. Lett.}, 2007, \textbf{98},
  150603\relax
\mciteBstWouldAddEndPuncttrue
\mciteSetBstMidEndSepPunct{\mcitedefaultmidpunct}
{\mcitedefaultendpunct}{\mcitedefaultseppunct}\relax
\EndOfBibitem
\bibitem[Yang and Ripoll(2011)]{yang:11}
M.~Yang and M.~Ripoll, \emph{Phys. Rev. E}, 2011, \textbf{84}, 061401\relax
\mciteBstWouldAddEndPuncttrue
\mciteSetBstMidEndSepPunct{\mcitedefaultmidpunct}
{\mcitedefaultendpunct}{\mcitedefaultseppunct}\relax
\EndOfBibitem
\bibitem[Earl \emph{et~al.}(2007)Earl, Pooley, Ryder, Bredberg, and
  Yeomans]{earl:07}
D.~J. Earl, C.~M. Pooley, J.~F. Ryder, I.~Bredberg and J.~M. Yeomans, \emph{J.
  Chem. Phys.}, 2007, \textbf{126}, 064703\relax
\mciteBstWouldAddEndPuncttrue
\mciteSetBstMidEndSepPunct{\mcitedefaultmidpunct}
{\mcitedefaultendpunct}{\mcitedefaultseppunct}\relax
\EndOfBibitem
\bibitem[Elgeti \emph{et~al.}(2010)Elgeti, Kaupp, and Gompper]{elge:10}
J.~Elgeti, U.~B. Kaupp and G.~Gompper, \emph{Biophys. J.}, 2010, \textbf{99},
  1018--1026\relax
\mciteBstWouldAddEndPuncttrue
\mciteSetBstMidEndSepPunct{\mcitedefaultmidpunct}
{\mcitedefaultendpunct}{\mcitedefaultseppunct}\relax
\EndOfBibitem
\bibitem[Elgeti and Gompper(2013)]{elge:13}
J.~Elgeti and G.~Gompper, \emph{Proc. Natl. Acad. Sci. USA}, 2013,
  \textbf{110}, 4470\relax
\mciteBstWouldAddEndPuncttrue
\mciteSetBstMidEndSepPunct{\mcitedefaultmidpunct}
{\mcitedefaultendpunct}{\mcitedefaultseppunct}\relax
\EndOfBibitem
\bibitem[Theers and Winkler(2013)]{thee:13}
M.~Theers and R.~G. Winkler, \emph{Phys. Rev. E}, 2013, \textbf{88},
  023012\relax
\mciteBstWouldAddEndPuncttrue
\mciteSetBstMidEndSepPunct{\mcitedefaultmidpunct}
{\mcitedefaultendpunct}{\mcitedefaultseppunct}\relax
\EndOfBibitem
\bibitem[Hu \emph{et~al.}(2015)Hu, Yang, Gompper, and Winkler]{hu:15.1}
J.~Hu, M.~Yang, G.~Gompper and R.~G. Winkler, \emph{Soft Matter}, 2015,
  \textbf{11}, 7843\relax
\mciteBstWouldAddEndPuncttrue
\mciteSetBstMidEndSepPunct{\mcitedefaultmidpunct}
{\mcitedefaultendpunct}{\mcitedefaultseppunct}\relax
\EndOfBibitem
\bibitem[Happel and Brenner(1983)]{happ:83}
J.~Happel and H.~Brenner, \emph{Low {Reynolds} number hydrodynamics: with
  special applications to particulate media}, Springer Science \& Business
  Media, 1983\relax
\mciteBstWouldAddEndPuncttrue
\mciteSetBstMidEndSepPunct{\mcitedefaultmidpunct}
{\mcitedefaultendpunct}{\mcitedefaultseppunct}\relax
\EndOfBibitem
\bibitem[Dassios and Vafeas(2008)]{dass:08}
G.~Dassios and P.~Vafeas, \emph{Physics Research International}, 2008,
  \textbf{2008}, e135289\relax
\mciteBstWouldAddEndPuncttrue
\mciteSetBstMidEndSepPunct{\mcitedefaultmidpunct}
{\mcitedefaultendpunct}{\mcitedefaultseppunct}\relax
\EndOfBibitem
\bibitem[Leshansky \emph{et~al.}(2007)Leshansky, Kenneth, Gat, and
  Avron]{lesh:07}
A.~M. Leshansky, O.~Kenneth, O.~Gat and J.~E. Avron, \emph{New Journal of
  Physics}, 2007, \textbf{9}, 145--145\relax
\mciteBstWouldAddEndPuncttrue
\mciteSetBstMidEndSepPunct{\mcitedefaultmidpunct}
{\mcitedefaultendpunct}{\mcitedefaultseppunct}\relax
\EndOfBibitem
\bibitem[Allahyarov and Gompper(2002)]{alla:02}
E.~Allahyarov and G.~Gompper, \emph{Phys. Rev. E}, 2002, \textbf{66},
  036702\relax
\mciteBstWouldAddEndPuncttrue
\mciteSetBstMidEndSepPunct{\mcitedefaultmidpunct}
{\mcitedefaultendpunct}{\mcitedefaultseppunct}\relax
\EndOfBibitem
\bibitem[Noguchi \emph{et~al.}(2007)Noguchi, Kikuchi, and Gompper]{nogu:07}
H.~Noguchi, N.~Kikuchi and G.~Gompper, \emph{EPL (Europhysics Letters)}, 2007,
  \textbf{78}, 10005\relax
\mciteBstWouldAddEndPuncttrue
\mciteSetBstMidEndSepPunct{\mcitedefaultmidpunct}
{\mcitedefaultendpunct}{\mcitedefaultseppunct}\relax
\EndOfBibitem
\bibitem[Noguchi and Gompper(2008)]{nogu:08}
H.~Noguchi and G.~Gompper, \emph{Phys. Rev. E}, 2008, \textbf{78}, 016706\relax
\mciteBstWouldAddEndPuncttrue
\mciteSetBstMidEndSepPunct{\mcitedefaultmidpunct}
{\mcitedefaultendpunct}{\mcitedefaultseppunct}\relax
\EndOfBibitem
\bibitem[Theers and Winkler(2015)]{thee:15}
M.~Theers and R.~G. Winkler, \emph{Phys. Rev. E}, 2015, \textbf{91},
  033309\relax
\mciteBstWouldAddEndPuncttrue
\mciteSetBstMidEndSepPunct{\mcitedefaultmidpunct}
{\mcitedefaultendpunct}{\mcitedefaultseppunct}\relax
\EndOfBibitem
\bibitem[Ihle and Kroll(2001)]{ihle:01}
T.~Ihle and D.~M. Kroll, \emph{Phys. Rev. E}, 2001, \textbf{63}, 020201\relax
\mciteBstWouldAddEndPuncttrue
\mciteSetBstMidEndSepPunct{\mcitedefaultmidpunct}
{\mcitedefaultendpunct}{\mcitedefaultseppunct}\relax
\EndOfBibitem
\bibitem[Ihle and Kroll(2003)]{ihle:03}
T.~Ihle and D.~M. Kroll, \emph{Phys. Rev. E}, 2003, \textbf{67}, 066705\relax
\mciteBstWouldAddEndPuncttrue
\mciteSetBstMidEndSepPunct{\mcitedefaultmidpunct}
{\mcitedefaultendpunct}{\mcitedefaultseppunct}\relax
\EndOfBibitem
\bibitem[Huang \emph{et~al.}(2010)Huang, Chatterji, Sutmann, Gompper, and
  Winkler]{huan:10}
C.~C. Huang, A.~Chatterji, G.~Sutmann, G.~Gompper and R.~G. Winkler, \emph{J.
  Comput. Phys.}, 2010, \textbf{229}, 168--177\relax
\mciteBstWouldAddEndPuncttrue
\mciteSetBstMidEndSepPunct{\mcitedefaultmidpunct}
{\mcitedefaultendpunct}{\mcitedefaultseppunct}\relax
\EndOfBibitem
\bibitem[Huang \emph{et~al.}(2015)Huang, Varghese, Gompper, and
  Winkler]{huan:15}
C.-C. Huang, A.~Varghese, G.~Gompper and R.~G. Winkler, \emph{Phys. Rev. E},
  2015, \textbf{91}, 013310\relax
\mciteBstWouldAddEndPuncttrue
\mciteSetBstMidEndSepPunct{\mcitedefaultmidpunct}
{\mcitedefaultendpunct}{\mcitedefaultseppunct}\relax
\EndOfBibitem
\bibitem[Westphal \emph{et~al.}(2014)Westphal, Singh, Huang, Gompper, and
  Winkler]{west:14}
E.~Westphal, S.~P. Singh, C.~C. Huang, G.~Gompper and R.~G. Winkler,
  \emph{Comput. Phys. Commun.}, 2014, \textbf{185}, 495--503\relax
\mciteBstWouldAddEndPuncttrue
\mciteSetBstMidEndSepPunct{\mcitedefaultmidpunct}
{\mcitedefaultendpunct}{\mcitedefaultseppunct}\relax
\EndOfBibitem
\bibitem[Allen and Tildesley(1987)]{alle:87}
M.~P. Allen and D.~J. Tildesley, \emph{Computer Simulation of Liquids},
  Clarendon Press, Oxford, 1987\relax
\mciteBstWouldAddEndPuncttrue
\mciteSetBstMidEndSepPunct{\mcitedefaultmidpunct}
{\mcitedefaultendpunct}{\mcitedefaultseppunct}\relax
\EndOfBibitem
\bibitem[Padding and Louis(2006)]{padd:06}
J.~T. Padding and A.~A. Louis, \emph{Phys. Rev. E}, 2006, \textbf{74},
  031402\relax
\mciteBstWouldAddEndPuncttrue
\mciteSetBstMidEndSepPunct{\mcitedefaultmidpunct}
{\mcitedefaultendpunct}{\mcitedefaultseppunct}\relax
\EndOfBibitem
\bibitem[Downton and Stark(2009)]{down:09}
M.~T. Downton and H.~Stark, \emph{J. Phys.: Condens. Matter}, 2009,
  \textbf{21}, 204101\relax
\mciteBstWouldAddEndPuncttrue
\mciteSetBstMidEndSepPunct{\mcitedefaultmidpunct}
{\mcitedefaultendpunct}{\mcitedefaultseppunct}\relax
\EndOfBibitem
\bibitem[Lamura \emph{et~al.}(2001)Lamura, Gompper, Ihle, and Kroll]{lamu:01}
A.~Lamura, G.~Gompper, T.~Ihle and D.~M. Kroll, \emph{EPL (Europhysics
  Letters)}, 2001, \textbf{56}, 319\relax
\mciteBstWouldAddEndPuncttrue
\mciteSetBstMidEndSepPunct{\mcitedefaultmidpunct}
{\mcitedefaultendpunct}{\mcitedefaultseppunct}\relax
\EndOfBibitem
\bibitem[Omelyan(1998)]{omel:98}
I.~P. Omelyan, \emph{Phys. Rev. E}, 1998, \textbf{58}, 1169--1172\relax
\mciteBstWouldAddEndPuncttrue
\mciteSetBstMidEndSepPunct{\mcitedefaultmidpunct}
{\mcitedefaultendpunct}{\mcitedefaultseppunct}\relax
\EndOfBibitem
\bibitem[Favro(1960)]{favr:60}
L.~D. Favro, \emph{Phys. Rev.}, 1960, \textbf{119}, 53--62\relax
\mciteBstWouldAddEndPuncttrue
\mciteSetBstMidEndSepPunct{\mcitedefaultmidpunct}
{\mcitedefaultendpunct}{\mcitedefaultseppunct}\relax
\EndOfBibitem
\bibitem[Kim and Karrila(2013)]{kim:13}
S.~Kim and S.~J. Karrila, \emph{Microhydrodynamics: {Principles} and {Selected}
  {Applications}}, Butterworth-Heinemann, 2013\relax
\mciteBstWouldAddEndPuncttrue
\mciteSetBstMidEndSepPunct{\mcitedefaultmidpunct}
{\mcitedefaultendpunct}{\mcitedefaultseppunct}\relax
\EndOfBibitem
\bibitem[G{\"o}tze and Gompper(2010)]{goet:10}
I.~O. G{\"o}tze and G.~Gompper, \emph{Phys. Rev. E}, 2010, \textbf{82},
  041921\relax
\mciteBstWouldAddEndPuncttrue
\mciteSetBstMidEndSepPunct{\mcitedefaultmidpunct}
{\mcitedefaultendpunct}{\mcitedefaultseppunct}\relax
\EndOfBibitem
\bibitem[Felderhof(2016)]{feld:16}
B.~U. Felderhof, \emph{arXiv:1603.08574 [physics]}, 2016\relax
\mciteBstWouldAddEndPuncttrue
\mciteSetBstMidEndSepPunct{\mcitedefaultmidpunct}
{\mcitedefaultendpunct}{\mcitedefaultseppunct}\relax
\EndOfBibitem
\bibitem[Wang and Guo(1989)]{wang:89}
Z.~X. Wang and D.~R. Guo, \emph{Special {Functions}}, World Scientific,
  1989\relax
\mciteBstWouldAddEndPuncttrue
\mciteSetBstMidEndSepPunct{\mcitedefaultmidpunct}
{\mcitedefaultendpunct}{\mcitedefaultseppunct}\relax
\EndOfBibitem
\bibitem[Paramonov and Yaliraki(2005)]{para:05}
L.~Paramonov and S.~N. Yaliraki, \emph{J. Chem. Phys.}, 2005, \textbf{123},
  194111\relax
\mciteBstWouldAddEndPuncttrue
\mciteSetBstMidEndSepPunct{\mcitedefaultmidpunct}
{\mcitedefaultendpunct}{\mcitedefaultseppunct}\relax
\EndOfBibitem
\bibitem[Press(2007)]{pres:07}
W.~H. Press, \emph{Numerical {Recipes} 3rd {Edition}: {The} {Art} of
  {Scientific} {Computing}}, Cambridge University Press, 2007\relax
\mciteBstWouldAddEndPuncttrue
\mciteSetBstMidEndSepPunct{\mcitedefaultmidpunct}
{\mcitedefaultendpunct}{\mcitedefaultseppunct}\relax
\EndOfBibitem
\end{mcitethebibliography}
\providecommand*{\mcitethebibliography}{\thebibliography}
\csname @ifundefined\endcsname{endmcitethebibliography}
{\let\endmcitethebibliography\endthebibliography}{}

}

\end{document}